\documentclass[useAMS,usenatbib,babel]{mn2e}

\usepackage[english,english]{babel}
\usepackage{amsmath}
\usepackage{amssymb,amsfonts,textcomp}
\usepackage{array}
\usepackage{supertabular}
\usepackage{hhline}
\usepackage{hyperref}  
\hypersetup{colorlinks=true, linkcolor=black, citecolor=black, filecolor=black, urlcolor=black}

\usepackage{flushend}

\usepackage{textcomp}
\usepackage{times}

\usepackage{graphicx}

\usepackage{dcolumn}
\usepackage{bm}
\usepackage{comment}
\usepackage{color}

\DeclareGraphicsExtensions{.jpg,.mps,.pdf,.png}

\newcommand{\mP}{{\cal P}}

\newcommand{\dd}{\hbox{d}}
\newcommand{\ii}{\hbox{i}}

\newcommand{\hrho}{{\hat \rho}}
\newcommand{\hs}{{\hat{s}}}

\newcommand{\lin}{{\rm lin}}

\DeclareMathOperator*{\argmax}{argmax}

\begin{document}

\author[C. Uhlemann,  S. Codis, J. Kim, C. Pichon et al. ]{
\parbox[t]{\textwidth}
{C. Uhlemann$^{1}$, S. Codis$^2$, J. Kim$^3$, C. Pichon$^{3,4}$, F.~Bernardeau$^{4,5}$, \\
D.~Pogosyan$^{4,6}$, C. Park$^{3}$ and B. L'Huillier$^{3}$}
\vspace*{6pt}\\
\noindent 
$^{1}$ Institute for Theoretical Physics, Utrecht University, Princetonplein 5, 3584 CC Utrecht, The Netherlands\\
$^{2}$ Canadian Institute for Theoretical Astrophysics, University of Toronto, 60 St. George Street, Toronto, ON M5S 3H8, Canada\\
$^{3}$ Korea Institute for Advanced Study (KIAS), 85 Hoegiro, Dongdaemun-gu, Seoul, 02455, Republic of Korea\\
$^{4}$ CNRS \& UPMC, UMR 7095, Institut d'Astrophysique de Paris, F-75014, Paris, France\\
$^{5}$ CNRS \& CEA, UMR 3681, Institut de Physique Th\'eorique, F-91191 Gif-sur-Yvette, France\\
$^{6}$ Department of Physics, University of Alberta, 412 Avadh Bhatia Physics Laboratory, Edmonton, Alberta, T6G 2J1, Canada  \\
}
\title[Non-linear two-point statistics of cosmic densities]{
Beyond Kaiser bias: mildly non-linear two-point statistics\\ 
of densities in distant spheres
}

\maketitle
\begin{abstract}
{
Simple parameter-free analytic bias functions for the two-point correlation of densities in spheres at large separation are presented. These bias functions generalize the so-called Kaiser bias to the mildly non-linear regime for arbitrary density contrasts as $b(\rho)-b(1) \propto{(1-\rho^{-13/21})\rho^{{1+n}/{3}}}$ with $b(1)=-4/21-n/3$ for a power-law initial spectrum with index $n$ . The derivation is carried out in the context of large deviation statistics while relying on the spherical collapse model. A logarithmic transformation provides a saddle approximation which is valid for the whole range of densities and shown to be accurate against the 30 Gpc cube state-of-the-art  Horizon Run~4 simulation. 
Special configurations of two concentric spheres that allow to identify peaks are employed to obtain the conditional bias and a proxy to  BBKS extrema correlation functions. These analytic bias functions
 should be used jointly with extended perturbation theory to predict two-point clustering statistics as they capture the non-linear regime of structure formation at the percent level down to scales of about $10$ Mpc$/h$ at redshift $0$. Conversely, the joint statistics also provide us with optimal dark matter two-point correlation estimates which can be applied either universally to all spheres or to a restricted set of biased (over- or underdense) pairs.  Based on a simple fiducial survey, this  estimator is shown to perform five times better than usual two-point function estimators. 
Extracting more information from correlations of different types of objects should prove essential in the context of upcoming surveys like 
Euclid, DESI, PFS or LSST.
}
\end{abstract}

\section{Introduction}

The  large-scale structure of the Universe  puts very tight constraints on  cosmological models. Deep spectroscopic surveys, like Euclid \citep{Euclid}, DESI \citep{DESI}, PFS \citep{2014PASJ...66R...1T} or LSST \citep{lsst}, will  allow astronomers to study the details of structure formation at different epochs, hence to probe cosmic acceleration. 
Yet, in order to reach the expected  precision on the equation of state of dark energy, astronomers  must address  the following challenges: non-linear gravitational evolution \citep{Bernardeau02}, redshift space distortions \citep{Kaiser87,tns}, bias \citep{Kaiser84,dekel87}, intrinsic alignments \citep{2015SSRv..193...67K} and baryonic physics \citep{2015JCAP...12..049S}. 
  
In this context, two-point clustering  has generated a lot of interest \cite[e.g.][and references therein]{halo-model},  as it allows one to investigate how the densest regions of space -- where dark halos usually reside -- are clustered,
 which in turn  sheds light on  the so-called biasing  between dark matter and halos:
as halos correspond to  peaks of the density field, they  are not  a  fair tracer of that field. 
\cite{Kaiser84} showed that 
in the high contrast, $\delta/\sigma \gg 1$, large separation limit, the correlation function, $\xi_{>\delta/\sigma}$, of peaks lying above this threshold reads
\begin{equation}
\xi_{>\delta/\sigma}\approx \frac{1}{\sigma^2} \left(\frac{\delta}{\sigma}\right)^{2}\xi,
\end{equation}
so that the correlation function of high density regions decreases more slowly than the density field correlation function, $\xi$, with an amplification factor or {\sl bias} that is proportional to the threshold squared. This analysis can also be restricted to the peaks of the density field above a given threshold following the seminal papers by \cite{BBKS} (hereafter BBKS) and \cite{Regos95}.
For two point functions,  the non-linear regime increases the number of   modes  used
to better constrain cosmological parameters. 
Of particular (partially theoretically unexplored) interest is the possibility of computing conditional two-point correlations, e.g. two-point correlation between regions that have specific densities,  so as to provide more robust estimates of the large-distance two-point correlation.

It has been argued \citep{Bernardeau14,Bernardeau15,Uhlemann16,Codis16b} that the statistics of cosmic densities in concentric spheres can leverage cosmic parameters 
competitively, as the corresponding spherical symmetry allows for analytical predictions in the mildly non-linear regime, beyond 
what is commonly achievable via other statistics in the context of perturbation theory. Indeed, the zero variance limit  of the cumulant generating functions
yields estimates of the joint density probability distribution function (PDF hereafter) which seems to match simulations in the regime of variances of order unity
  \citep{1989A&A...220....1B,1992ApJ...390L..61B,1993ApJ...412L...9J,2002A&A...382..412V,Bernardeau14,Bernardeau15}. This success was shown to originate from a regime of large deviations at play in the mildly non-linear evolution of the large-scale structure \citep{LDPinLSS}.

  The aim of this paper is to show that the spherically-symmetric framework which led to surprisingly accurate predictions for one-point statistics also accommodates, in the large separation limit, analytic
 estimates of the two-point statistics and in particular of the bias factor  associated with  imposed constraints within concentric cosmic densities.
Recently,
 \citeauthor{Codis16a} (\citeyear{Codis16a}a) studied the two-point statistics of the density within concentric spheres, whose redshift evolution was shown to be accurately predicted by large-deviations theory in the mildly non-linear regime, but relied on numerical integration of highly oscillating complex functions and  was therefore subject to possibly significant numerical errors, in particular for large densities.
Since \cite{Uhlemann16} showed that very accurate analytic approximations could be found for one-point statistics by using a logarithmic transform of the density field and performing a saddle-point approximation, we propose in this paper to extend the use of the logarithmic transform to two-point statistics. 

It was shown in \citeauthor{Codis16b} (\citeyear{Codis16b}b) that 
the one-point PDF can be fully predicted, modulo one parameter, the variance
of the density field, which is the driving parameter of the theory, leading to  
two options:  i) higher order perturbation theory 
can be used to predict the value of this variance as a function of scale and redshift in order to recover the full PDF or ii) this one parameter model can be used to build optimal likelihood estimators for the variance based on the measurement of densities in spheres. 
Conversely, in the present paper, modulo the unknown underlying two-point correlation function of the dark matter density field, we will show that the same large-deviations formalism provides us 
with the full statistics of  the two-point PDF of the density within concentric spheres separated by a distance $r_e$.
Once again, one can i) rely on perturbation theory to predict the underlying dark matter  correlation function  \citep[e.g.][]{Taruya2012},  
or ii) build, from the present theory, optimal estimators for the dark matter correlation function to be applied to measured density in separated spheres.

\begin{figure}
\centering\includegraphics[width=0.65\columnwidth]{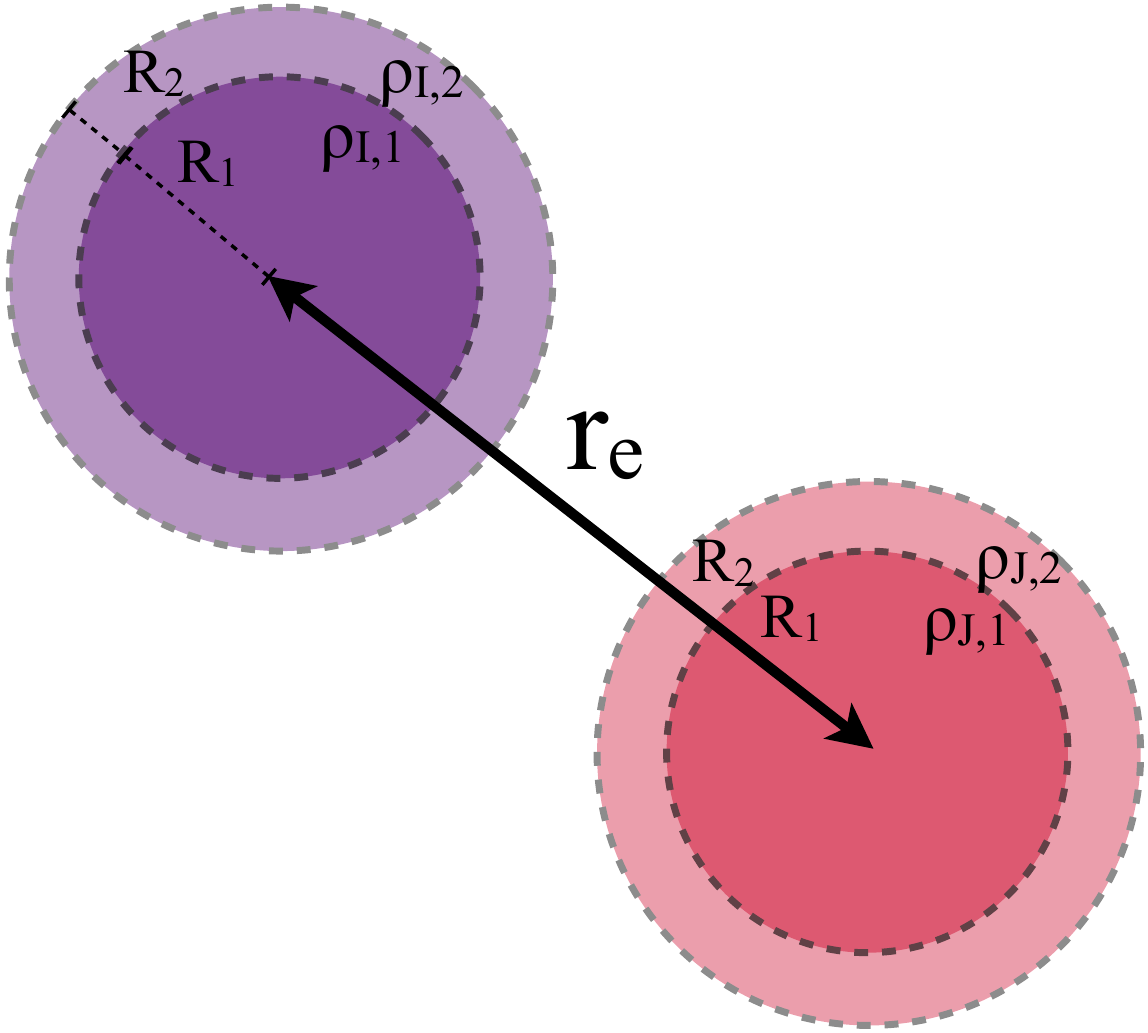}  
   \caption{The two-point configuration  with two concentric cells of radii $R_{1}$ and $R_{2}$ in one location (purple) and two other concentric cells of radii $R_{1}$ and $R_{2}$ in another location (red) separated by a distance
   $r_e\gg R_2$.   } 
   \label{fig:configuration-nplusncell}
\end{figure}

 In this paper, following \cite{1996A&A...312...11B}, the focus will be on predicting analytically the density two-point statistics for configurations shown in Figure~\ref{fig:configuration-nplusncell} and
specifically the corresponding bias functions (the aforementioned density-dependent scalings of the two-point correlation). We will in particular consider the density field smoothed at two different scales in two concentric spheres which can be turned into an inner density and a slope (difference of density between the two spheres). This will allow us to focus on
the conditional density-given-slope bias 
as a quasi-linear proxy for the BBKS peak correlation function.
 These bias functions generalize the so-called Kaiser linear bias in the mildly non-linear regime for large separations and arbitrary density contrasts.
Hence they provide alternative ways of using  gravitational clustering to probe our cosmological model, in particular using specific regions of space (underdense/overdense, small/big slope, etc).
Leveraging conditionals on the value of the density at the legs of the correlations will allow for a more robust estimate of the two-point correlation function.
We will illustrate on a fiducial experiment how the present formalism can be used to estimate optimally the underlying top-hat filtered correlation function. 

This paper is organized as follows. Section~\ref{sec:LDPS} presents briefly the implementation of a  large deviation principle on 
the joint statistics of concentric cells based on a saddle point approximation.
Section~\ref{sec:validation}  compares these analytic predictions to the state-of-the-art dark matter simulation Horizon~Run~4 (HR4).
Section~\ref{sec:recoverxi} demonstrates how to measure optimally  the dark matter correlation function on a given survey.
 Finally, Section~\ref{sec:conclusion} wraps up.
Appendix~\ref{app:LSSfast} shortly describes the accompanying package  {\tt LSSFast} for the evaluation of the one-cell PDF and the bias functions.
Appendix~\ref{app:LDP} reviews the formalism of large deviations relevant to obtain the density PDF for concentric spheres and the joint PDF at large separations. Appendix~\ref{app:biasPT} provides a description of bias functions in the Gaussian and weakly non-Gaussian regime based on perturbation theory. 
Appendix~\ref{app:figures} provides a validation of HR4 at redshift $z=4$ together with extended results for redshift $z=0$.

 \section{Large deviations and separation}
 \label{sec:LDPS}
 
 Appendix~\ref{app:LDP}  presents rapidly the general formalism for deriving the PDF and the bias functions from the large deviation principle. 
 Let us in the main text   focus on presenting 
directly the corresponding fully analytical predictions, relying on the so-called saddle point approximation. 
For that purpose,  we shall see that all that is needed is the so-called decay-rate function (controlling the exponential decay of the probablity distribution with the density at leading order),
the key  quantity  that connects the (Gaussian) initial  to the non-Gaussian final distribution  brought about by non-linear  gravitational clustering.

In this section we will determine the PDF of densities in {\sl distant spheres} based on large deviation statistics and the saddle point approximation. 
This  approximation has proven successful already to predict  the one- and two-cell PDF for densities in {\sl concentric spheres} in \cite{Uhlemann16}. After introducing the two-point PDF and the bias functions in Section~\ref{sec:defbias} and the spherical collapse model as the mapping for large deviation statistics in Section~\ref{sec:SC}, we shortly review results for the one-point PDF in Section~\ref{sec:onepointPDF} and finally extend them to the two-point PDF of densities at large separations, hence the bias functions in Section~\ref{sec:biaslarge}.\footnote{Note that, by constructing PDFs for densities in a larger number of concentric spheres in special configurations, such as three spheres, with a central density 
within the inner sphere and a second density within a  given outer shell, one could build correlators for arbitrary separation. This will be the topic of upcoming investigations.}  The joint knowledge of the PDF and the bias functions enables us to compute constrained bias functions which are of particular interest for the clustering of over- and underdense regions.  
  
 \subsection{Definition of the bias functions}
 \label{sec:defbias}

Let us consider two sets of $N$ concentric spheres separated by a distance $r_e$ and define the corresponding densities $\{\rho_{k}\}_{1\leq k\leq N}$ and $\{\rho'_{k}\}_{1\leq k\leq N}$. Following \cite{1996A&A...312...11B} and \citeauthor{Codis16a} (\citeyear{Codis16a}a), the joint PDF of these densities, $\mP(\{\rho_{k}\},\{\rho_{k}'\};r_e)$, can be predicted from the one-point PDFs $\mP(\{\rho_{k}\})$, $\mP(\{\rho_{k}'\})$ and the correlation function of densities in spheres at finite separation 
\begin{subequations}
\begin{equation}
 \mP(\{\rho_{k}\},\{\rho'_{k}\};r_{e})= \mP(\{\rho_{k}\})\mP(\{\rho'_{k}\})\left[1+\xi_{\circ}(\{\rho_{k}\},\{\rho_{k}'\};r_{e})\right] \\ \label{eq:jointPDF0}
\end{equation}
where the sphere correlation function, $\xi_{\circ}(\{\rho_{k}\},\{\rho_{k}'\};r_{e})$, at large separations $r_e\gg R_k$ can be related to the underlying (unbiased) top-hat smoothed dark matter two-point correlation function $\xi(r_{e})$ via the effective bias functions $b(\{\rho_{k}\})$ and $b(\{\rho'_{k}\})$:
\begin{equation}
\xi_{\circ}(\{\rho_{k}\},\{\rho_{k}'\};r_{e})  =\xi(r_{e})b(\{\rho_{k}\})b(\{\rho_k '\})\,. \label{eq:jointPDF}
\end{equation}
\end{subequations}
{This is the count-in-cell analog of the so-called peak background split and defines the bias factor $b(\{\hrho_k\})$.} Physically, this bias encodes the mean density in a sphere given that the densities in spheres of radii $\{R_k\}$ at large separation $r_e$ are $\{\hat\rho_k\}$
\begin{align}
\label{eq:brho-N-cell}
 1+b(\{\hrho_k\})\xi(r_{e}) &= \langle\rho'|\{\hrho_k\}\, ;r_e\rangle\,,
\\
\notag
& 
=\frac{\textstyle\int_{0}^{\infty}\dd \rho ' \mP(\{\hrho_k\},\rho ';r_{e}) \rho '}
{\mP(\{\hrho_k\})} \,,
\end{align}
which follows from \eqref{eq:jointPDF0} by integration and normalization of the PDFs. In the following, we will describe how this bias function, and hence the joint density PDF at large separation can be predicted analytically.

\subsection{Large deviation statistics with spherical collapse}
\label{sec:SC}
When considering a highly symmetric observable such as the density in spheres, one can argue that the most likely dynamics (amongst all possible mappings between the initial and final density field) is the one respecting the symmetry \citep{2002A&A...382..412V}.\footnote{{This is a result of the so-called contraction principle in the context of large deviation theory as explained in \cite{LDPinLSS},
which formalizes the idea that amongst all unlikely fates (in the tail of the PDF) the least unlikely one (spherical collapse) dominates.}} For spherical symmetry, one can then take advantage of the fact that non-linear solutions to the gravitational dynamics are known explicitly in terms of the spherical collapse model.

Let us denote $\zeta_{\rm SC}(\tau)$ the non-linear transform of an initial fluctuation with linear density contrast, $\tau$, in a sphere of radius $r$, to the final density $\rho$ (in units of the average density) in a sphere of radius $R$ according to the spherical collapse model
\begin{equation}
\rho=\zeta_{\rm SC}(\tau)\,, 
\quad {\rm with}
\quad
\rho R^{3}= r^{3}\,, 
\label{eq:rho2tau}
\end{equation}
where the initial and final radii are connected through mass conservation. An explicit possible fit for $\rho_{\rm SC}(\tau)$ is given by
\begin{equation}
\rho_{\rm SC}(\tau)=(1-\tau/\nu)^{-\nu}\,, \label{eq:spherical-collapse}
\end{equation}
where $\nu$ can be adjusted to the actual values of the cosmological parameters ($\nu=21/13$ provides 
a good description of the spherical dynamics for an Einstein-de Sitter background for the range of $\tau$
values of interest).

Thanks to this analytic spherical collapse model, the one-point PDF and bias functions of cosmic densities in concentric spheres, brought about by  non-linear gravitational evolution, can be predicted explicitly from the given (Gaussian) initial conditions.

\subsection{The density PDF in the large deviation regime}
\label{sec:onepointPDF}

\subsubsection{PDF and decay-rate function for an initial Gaussian field}
The principles of large deviation statistics yields a formula for the PDF of finding a certain density in a sphere given the initial conditions. The decay-rate function encodes the exponential decay of the PDF. For Gaussian initial conditions, which we assume here,
\begin{align}
 \label{eq:saddlePDFlogN-cell-ini}
\mP_{\{r_k\}}^{\rm{ini}}(\{\tau_k\})&= \sqrt{\det\left[\frac{\partial^{2}\Psi^{\rm ini}_{\{r_k\}}}{\partial \tau_{i}\partial \tau_{j}}\right]} \frac{ \exp\left[-\Psi^{\rm ini}_{\{r_k\}}(\{\tau_k\})\right]}{ (2 \pi)^{N/2}} 
\,,
 \end{align}
  the  initial decay-rate function is  given by the usual quadratic form in the initial density contrasts $\tau_k$
\begin{align}
\Psi^{\rm ini}_{\{r_k\}}(\{ \tau_{k}\})=\frac{1}{2}\sum_{i,j}\Xi_{ij}(\{ r_k \})\,\tau_{i}\tau_{j}\,,
\label{PsiDefIni}
\end{align}
where $\Xi_{ij}$ is the inverse of the initial covariance matrix, $\sigma_{ij}^{2}=\sigma^{2}(R_{i},R_{j})$, encoding all dependency with respect to the initial power spectrum according to
\begin{align}
\label{eq:covmatrix}
\sigma^{2}_{ij}&=\int\frac{\dd^{3}k}{(2\pi)^{3}}P^{\lin }(k)W_{\rm 3D}(k R_{i})W_{\rm 3D}(k R_{j})\,,
\end{align}
where $W_{\rm 3D}$ is the Fourier transform of  the top-hat filter
\begin{equation}
W_{\rm 3D}(k)=\frac{3}{k^{2}}(\sin(k)/k-\cos(k))\,.
\end{equation}
Note that equation~\eqref{eq:saddlePDFlogN-cell-ini} is an unusual rewrite of a Gaussian distribution, emphasizing the central role of the 
rate function~\eqref{PsiDefIni}. This rate function 
has a straightforward explicit expression in terms of the underlying covariances hence the initial power spectrum. 

\subsubsection{Saddle-point PDF for an evolved non-Gaussian field}  

The final decay-rate function is obtained from re-expressing the initial decay-rate function in terms of the final densities
\begin{align}
\Psi_{\{R_k\}}(\{ \rho_{k}\})=\frac{1}{2}\sum_{i,j}\Xi_{ij}(\{ R_k\rho_{k}^{1/3} \})\,\tau_{i}(\rho_{i})\tau_{j}(\rho_{j})\,,
\label{PsiDef}
\end{align}
using the spherical collapse mapping characterized by equation~\eqref{eq:spherical-collapse}. The previously known result for the PDF of densities in concentric spheres $\mP(\{\rho_k\})$ at one point is given by
\begin{align}
\label{eq:saddlePDFlogN-cell-rho}
\hskip -0.25cm \mP_{\{R_k\}}(\{\rho_k\})&= \sqrt{\!\det\!\left[\frac{\partial^{2}\Psi_{\{R_k\}}}{\partial \rho_{i}\partial \rho_{j}}\right]} \frac{ \exp\left[-\Psi_{\{R_k\}}(\{\rho_k\})\right]}{ (2 \pi)^{N/2}} ,
 \end{align}
 where $\Psi$ is given by equation~\eqref{PsiDef}.
The saddle point approximation provides a very good approximation to the exact result from large-deviations statistics, as discussed in  Appendix~\ref{app:LDP}, if the final decay-rate function is convex, i.e. $$\det\left[\frac{\partial^{2}\Psi_{\{R_k\}}}{\partial \rho_{i}\partial \rho_{j}}\right] {>}0 \,,$$ which simplifies to $\Psi''_R[\rho]>0$ for the one-cell case. However, as has been shown in \cite{Bernardeau14}, this condition is only fulfilled below a critical value $\rho_c$ where $\Psi_R''[\rho_c]=0$ in the one-cell case, and similarly inside a $N-1$ manifold for the $N$-cell case. The main point 
of \cite{Uhlemann16} is that this difficulty can be alleviated with an adequate change of variables such as the logarithmic density transform. The procedure is then to apply the saddle-point approximation to predict the PDF of the (logarithmically) mapped density field as 
\begin{subequations}
 \label{eq:saddlePDFN-cell}
 \begin{align}
 \label{eq:saddlePDFlogN-cell}
\mP_{\mu,\{R_k\}}(\{\mu_k\})&= \sqrt{\det\left[\frac{\partial^{2}\Psi_{\{R_k\}}}{\partial \mu_{i}\partial \mu_{j}}\right]} \frac{ \exp\left[-\Psi_{\{R_k\}}\right]}{ (2 \pi)^{N/2}} 
\,,
 \end{align}
 where the transformation $\{\rho_k\}\rightarrow\{\mu_k\}$ has to be chosen to ensure the convexity of the decay-rate function. This result can then easily be translated in the PDF of the density field via a change of variables
\begin{align}
\mP_{\{R_k\}}(\{\rho_k\}) &=\mP_{\mu,\{R_k\}}[\{\mu_k(\{\rho_i\})\}] \left|\det\left[\frac{\partial\mu_{i}}{\partial \rho_{j}}\right]\right|  \,,
\label{eq:Prhofrommu}
\end{align}
where the Hessian of the decay rate function $\Psi_{\{R_k\}}$ after a change of variables $\{\rho_k\}\rightarrow\{\mu_k\}$ is trivially given by
\begin{align}
\frac{\partial^{2}\Psi_{\{R_k\}}}{\partial \mu_{i}\partial \mu_{j}}=\frac{\partial\rho_k}{\partial\mu_i}\cdot \frac{\partial^{2}\Psi_{\{R_k\}}}{\partial \rho_{k}\partial \rho_{l}}\cdot \frac{\partial \rho_l}{\partial\mu_j} + \frac{\partial^2\rho_k}{\partial\mu_i\partial\mu_j}\cdot \frac{\partial\Psi_{\{R_k\}}}{\partial\rho_k} \,.\notag
\end{align}

\subsubsection{Ensuring normalization}
Equation~(\ref{eq:saddlePDFN-cell}) assumes that the mean of $\mu_{j}$ does not depend on the variance and vanishes. For a generic non-linear mapping, it will translate into a mean density which can deviate from one as $\sigma$ grows. In order to avoid this effect, one has to consider the shifted PDF 
\begin{equation}
\label{eq:saddlePDFN-cellnorm}
\hat\mP_{\mu,\{R_k\}}(\{\mu_k\})=\mP_{\mu,\{R_k\}}(\{\tilde\mu_k=\mu_k-\left\langle\mu_{k}\right\rangle\})\,,
\end{equation}
with the shifts $\left\langle\mu_{k}\right\rangle$ chosen such that the resulting mean densities are one $\langle\rho_i\rangle=1\,\forall i=1,\cdots,n$. Furthermore, since the saddle-point method yields only an approximation to the exact PDF, the PDF obtained from equation~\eqref{eq:saddlePDFN-cell} is not necessarily properly normalized. In practice, this can be accounted for by considering
\begin{equation}
\label{eq:saddlePDFN-cellnorm2}
\hat\mP_{R}(\{\rho_k\})=\mP_{R}(\{\rho_k\})/\langle1\rangle\,,
\end{equation}
with the shorthand notation $\langle 1\rangle= \prod_k \int_0^\infty \dd\rho_k\, \mP_{R}(\{\rho_k\})$.
\end{subequations}

\subsection{The bias functions at large separations}
\label{sec:biaslarge}
The saddle point approximation applies  also to the joint density PDF at large separation and hence the bias function defined in equation~\eqref{eq:brho-N-cell}.
Since initially, the field is Gaussian, the initial bias function is exactly given by the so-called Kaiser linear bias as described in section~\ref{sec:Kaiser}. The subsequent quasi-linear evolution can then be predicted by the large-deviations principle as will be shown in section~\ref{sec:bLDP}.

\subsubsection{Kaiser bias for an initial Gaussian field}
\label{sec:Kaiser}
Let us consider a set of density contrasts $\{\tau_{k}\}$ in concentric spheres of radii $R_{k} $ and the contrast
$\tau'_{1}$  in a sphere of radius $R_{1}$ at a distance $r_{e}$ away from the center of the concentric spheres. If the density field is Gaussian, the covariance matrix of $(\{\tau_{k}\},\tau'_{1})$ simply reads
\begin{equation}
\label{eq:Kaisercovmatrix}
C=\left(
\begin{array}{ccc}
 \sigma_{ij}^{2} & \xi_{i1}  \\
\xi_{1j}  & \sigma_{11}^{2} 
\end{array}
\right)\,,
\end{equation}
where we use the short hand notation $\sigma_{ij}^{2}=\sigma^{2}(R_{i},R_{j})$ and $\xi_{ij}=\xi(R_i,R_j;r_e)$ with 
\begin{align}
\xi_{ij}&=\int\frac{\dd^{3}k}{(2\pi)^{3}}P^{\lin }(k)W_{\rm 3D}(k R_{i})W_{\rm 3D}(k R_{j})\exp(\imath k r_{e}\cos\theta)\,. \notag
\end{align}
In the Gaussian case the bias function can be computed analytically, for example by diagonalizing the covariance matrix by a change of variables as shown in Appendix~\ref{app:biasPT}. The result for the one-cell density bias is
\begin{align}
\label{eq:bKaiser1cell}
b^{\rm G}(\tau)=\frac{\left\langle \tau'_{1}|\tau\right\rangle}{\xi_{11}}
=\frac{\tau}{\sigma_{11}^{2}}\,,\quad
\end{align}
which is proportional to the initial overdensity $\tau$ as expected from \citep{Kaiser84}. The $N$-cell density bias follows as
\begin{align}
\label{eq:bKaiser}
b^{\rm ini}(\{\tau_{k}\}) 
&= \sum_{i,j=1}^N \Xi_{ij}(r_i,r_j) \tau_i = \sum_{j=1}^N \frac{\partial\Psi^{\rm ini}_{\{r_k\}}(\{\tau_k\})}{\partial \tau_j}\,, 
\end{align}
if we assume that for large separations $r_{e}\gg R_k$ the cross-correlations are all approximately identical $\xi_{1i}\approx \xi_{11}\,\forall i$. 
In general, the Kaiser bias function is given in terms of the derivative of the decay-rate function of the initial PDF $\Psi_{\{r_k\}}(\{\tau_k\})=-\log\mP_{\{r_k\}}(\{\tau_k\})$ and hence the rate of decay of the PDF. This encodes the idea that unlikely configurations, corresponding to strongly positive or negative values of the initial density contrast $\tau$ are more biased. 

\subsubsection{Saddle-point bias for an evolved non-Gaussian field} 
\label{sec:bLDP}

The saddle point approximation of the bias function amounts to mapping the initial Kaiser bias function,
equation~\eqref{eq:bKaiser}, using the inverse spherical collapse dynamics from equation~\eqref{eq:spherical-collapse}
\begin{align}
\label{eq:bsaddle}
b(\{\rho_{k}\}) 
&= \sum_{i,j=1}^n \Xi_{ij}(R_i\rho_i^{1/3},R_j\rho_j^{1/3}) \tau_i(\rho_i)\,.
\end{align}
The spherical collapse can be shown to be the leading order contribution for the statistics of densities in distant sphere,  as was done for the one-point PDF\footnote{More precisely, the analytical asymptote of the bias function can be derived using a steepest descent method in equations~\eqref{eq:PDF} and \eqref{eq:defbiasPDF}, see Appendix~\ref{app:LDP}}.
This saddle point approximation is valid in the large separation regime and as long as the PDF of the density can be obtained via a saddle point approximation. In as much as the logarithmic transform significantly increases the region of applicability of the saddle point approximation for the PDF, it also yields analytical bias functions. Using this saddle point approximation we therefore extend the results of \citeauthor{Codis16b} (\citeyear{Codis16b}b) and present analytical predictions that do not require a numerical integration in the complex plane;
we also provide predictions for the joint and constrained biases based on analytical predictions of 2-cell quantities which were not easily accessible before.

\subsubsection{Ensuring normalization}

Because of the normalisation of the density and joint density PDFs and the definition of the sphere correlation function, the bias function must obey the following two relations
\begin{align}
\langle b(\{\rho_k\})\rangle = & \prod_k \int_{0}^{\infty} \!\!\dd \rho_{k}\, \hat\mP(\{\rho_{k}\})\,b(\{\rho_{k}\})=0\,, \label{eq:biasprop} \\
\notag \langle\rho_i b(\{\rho_k\})\rangle =& \prod_k \int_{0}^{\infty} \!\! \dd \rho_{k}\, \hat\mP(\{\rho_{k}\})\,b(\{\rho_{k}\}) \rho_{i}=1\,, \forall i=1,...,n\,. 
\end{align}
Because of these properties the bias function $b(\{\rho_k\})$ obtained from equation~\eqref{eq:bsaddle} still have to be normalized according to%
\begin{align}
\hat b(\{\rho_k\}) &= \frac{b(\{\rho_k\})- \langle b(\{\rho_k\})\rangle}{\frac{1}{n}\sum_{i=1}^n \left(\langle\rho_i  b(\{\rho_k\})\rangle- \langle b(\{\rho_k\})\rangle\right)}\,. \label{eq:biasnorm}
\end{align}
{This normalization procedure is necessary and can be understood easily: while the conditions~\eqref{eq:biasnorm} are trivially fulfilled for a purely Gaussian initial field with small variance and Kaiser bias \eqref{eq:bKaiser}, the bias will pick up corrections from non-Gaussianity via  gravitational collapse. Those corrections modify the value of the bias at average density (and hence the mean bias and mean density-weighted bias) and become manifest already in the very mildly non-Gaussian (and hence perturbative) regime as shown in Appendix~\ref{app:biasPT}. 
Those non-Gaussian corrections that affect the mean are not accounted for in the saddle point approximation used in this work. To correct for this effect, we choose to shift the non-perturbative result from spherical collapse dynamics at the end according to the nonzero mean bias and will show that it leads to accurate predictions that are robust to variances of order one. Furthermore, extrapolating the saddle point approximation to finite variances requires to adjust the absolute normalization connected to the mean density-weighted bias.}

 \section{Validation with the HR4 simulation}
 \label{sec:validation}

Let us now evaluate the simple analytical predictions for the density PDF in concentric spheres,
equation~\eqref{eq:saddlePDFlogN-cell-rho}, together  with the bias functions at large separations, equation~\eqref{eq:bsaddle}, and compare them to measurements in the HR4 simulation presented  in Section~\ref{ssec:HR4}. The estimators for the measurements of the bias functions are described in Section~\ref{ssec:measurebias} while the parametrization for the correlation function is given in Section~\ref{ssec:paramcov}.
For brevity, we will focus our comparison in the main text to redshift $z=0.7$, which is in the redshift range that is most interesting to current surveys, while results for  $z=0$ and a validation at high redshift $z=4$ are shown in Appendix~\ref{app:figures}.   
We will validate our analytical results in Section~\ref{ssec:PDF-conc} against the HR4 measurements, then
present conditional bias function in Section~\ref{ssec:biaslarge}
and 
discuss the modulation of the matter correlation function  induced by  biasing  in Section~\ref{ssec:corrfct}.

\subsection{The Horizon-Run 4 simulation}
\label{ssec:HR4}
\begin{figure}
\centering\includegraphics[width=0.95\columnwidth]{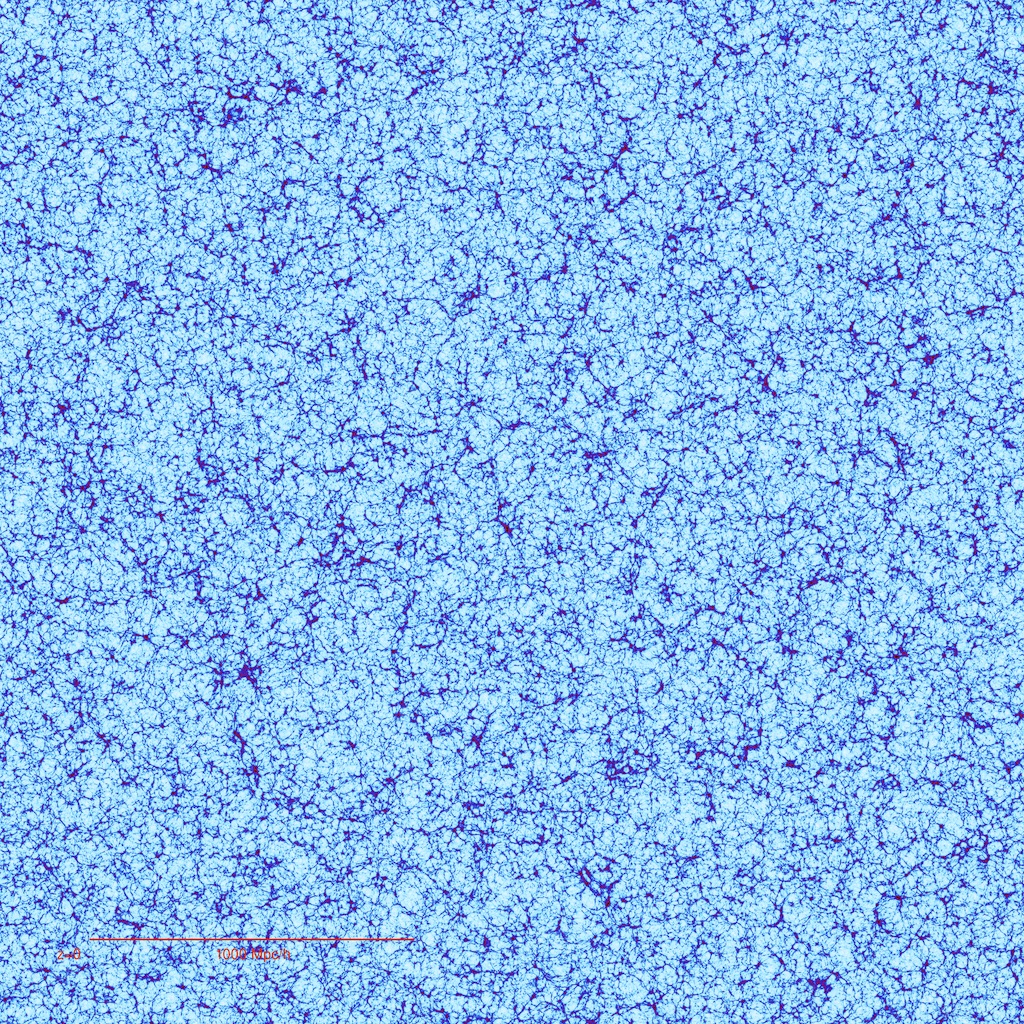}  
   \caption{A thin (1.5 Mpc$/h$) slice through the HR4 simulation at redshift 0.
   The simulation is 3150 Mpc$/h$ across sampled by $6300^3$ dark matter particles.} 
   \label{fig:HR4}
\end{figure}

The Horizon Run 4 (HR4) simulation \citep{2015JKAS...48..213K} is a state-of-the-art dark matter simulation modelling gravitational clustering  in a
Hubble-like volume. It 
 was run on the Tachyon-2 system at Korea Institute of Science and Technology Information to study of the nonlinear matter evolution using $N_p=6300^3$ particles in a cubic box with a side length of 3150 $h^{-1} {\rm Mpc}$. The particle mass is about $m_p\simeq 9.0\times 10^9 ~h^{-1} {\rm M_\odot}$. The adopted cosmology is comparable with the WMAP 5 year $\Lambda$CDM model. with $\sigma_8=0.794$ and  matter, baryonic matter, and dark energy density parameters of $\Omega_{m,0} = 0.26, \Omega_{b,0} = 0.044$, and $\Omega_{\Lambda,0} = 0.74$, respectively. The initial conditions were generated at $z_i=100$ according to the second-order Lagrangian Perturbation Theory (2LPT; Jenkins 2000). The gravitational force on each particle was calculated using the PM-Tree method by the GOTPM \citep{2004NewA....9..111D} down to $z=0$ in 2000 global time steps, see Figure~\ref{fig:HR4}.
The force resolution is 0.05$h^{-1}$Mpc, i.e. 1/10 of the mean particle separation.

To enhance the positional accuracy of particles, we adopt the shift vector rather than the position vector for particle information. In such a huge simulation like  HR4, the 32-bit floating-point accuracy has a round-off error around sub-\% level in terms of the mean particle separation, and a large time-step evolution may accumulate such errors and may affect the final matter distributions. Without requesting any more memory space, we devised a trick using the particle index as the Lagrangian position and the shift vector to calculate the particle position.  For more details about the simulation  and methods adopted by the GOTPM, see \cite{2015JKAS...48..213K}.

\subsection{Measuring the bias functions}
\label{ssec:measurebias}
To measure the bias functions we will make use of equations~\eqref{eq:brho-cross},~\eqref{eq:bsrho-cross},~\eqref{eq:brhoconstrs}~and~\eqref{eq:brhoconstrrho2}. In practice, we determine the values of the PDFs in bins of a certain width around the given value by counting the number of densities in a given bin using stepwise functions. 
The HR4 simulation density is estimated  via dark matter count-in-cell in $252^3$ cells separated by $12.5 {\rm Mpc}/h$.  
We use spheres of radii $R=10, 11, 12, 13, 14, 15\, {\rm Mpc}/h $ with a separation $r_e=37.5 {\rm Mpc}/h$ which is big enough to ensure that we are in the large separation regime, see  \citeauthor{Codis16a} (\citeyear{Codis16a}a).

\paragraph*{Correlation function}
We measure the sphere correlation function at distance $r_{e}$ according to
\begin{equation}
\label{eq:xi}
\hat \xi (r_{e})=\frac{\sum_{\rm I=1}^{\rm N_{t}}\sum_{j=1}^{6}\rho_{\rm I}\rho_{\alpha_{{\rm I},j}}}{6\mathrm{N_{t}}}-1.
\end{equation}
In practice, we count all pairs of spheres only once when computing $\hat \xi (r_{e})$ by only considering three neighbours for each sphere.

\paragraph*{Density bias}
The density bias is estimated using the cross-correlations of spheres with radius $R$ defined in equation~(\ref{eq:brho-cross}). More precisely, we compute a sum over each sphere $\rm I$ with density $\rho_{\rm I}$ and its 6 neighbours at distance $r_{e}$ labelled with the indices $\alpha_{{\rm I},j}$ for $1\leq j\leq 6$
\begin{equation}
\label{eq:effbrho}
\hat b (\hrho)=\frac 1 {\hat \xi}\!\left[\frac{\sum_{\rm I}\sum_{j=1}^{6}{\cal B}(|\rho_{\rm I}-\hrho|\leq\!\Delta\rho/2)\rho_{\alpha_{{\rm I},j}}
}{
6\sum_{\rm I}{\cal B}(|\rho_{\rm I}-\hrho|\leq\!\Delta\rho/2)}-1\right]\,,
\end{equation}
where ${\cal B}$ is a boolean function which evaluates to one if the density is in a bin centred on $\hrho$ with width $\Delta \rho =0.1$. Note that the density bias can also be measured using auto-correlations which has been shown to give results consistent with cross-correlations, see \citeauthor{Codis16a} (\citeyear{Codis16a}a).

\paragraph*{Joint density slope bias}
In order to measure the slope bias in the simulation, we consider the set of concentric spheres with radii $R_{1/2}$ such that $R_{2}-R_{1}=1 {\rm Mpc}/h $.
Following equation~(\ref{eq:bsrho-cross}), we compute again a sum over each set $\rm I$ and its 6 neighbouring sets at distance $r_{e}$ labelled with the indices $\alpha_{{\rm I},j}$ for $1\leq j\leq 6$
\begin{equation}
\nonumber
\hat b (\hrho,\hs)\!=\!\frac 1 {\hat \xi}\!\left[\!\frac{\sum_{\rm I,j}{\cal B}(|\rho_{\rm I}-\hrho|\!\leq\!\Delta\rho/2,|s_{\rm I}-\hs|\!\leq \!\Delta s/2)\rho_{\alpha_{{\rm I},j}}
}{6\sum_{\rm I}{\cal B}(|\rho_{\rm I}-\hrho|\!\leq\!\Delta\rho/2,|s_{\rm I}-\hs|\!\leq \!\Delta s/2)}\!-\!1\right]\!, \label{eq:effbrhos}
\end{equation}
${\cal B}$ is a boolean function which evaluates to one if the density is in a bin centred on $\hrho$ with width $\Delta \rho =0.2$ and the slope is in a bin centred on $\hs$ with width $\Delta s=0.02 R_1$. 

\paragraph*{Constrained density bias given environment}
From the joint bias of the density and slope one can also determine a constrained density bias given an environment either specified by 
a positive or negative slope 
\begin{equation}
\nonumber
\hat b (\hrho|s\!\lessgtr\!0)\!=\!\frac 1 {\hat \xi}\!\left[\!\frac{\sum_{\rm I,j}{\cal B}(|\rho_{\rm I}-\hrho_1|\!\leq\!\Delta\rho/2,s_{\rm I}\!\lessgtr\!0)\rho_{\alpha_{{\rm I},j}}
}{6\sum_{\rm I}{\cal B}(|\rho_{\rm I}-\hrho_1|\!\leq\!\Delta\rho/2,s_{\rm I}\!\lessgtr\!0)}\!-\!1\right]\,,\label{eq:effbrhoconstrs}
\end{equation}
or an over- or underdense shell $\rho_{12}\!=\!(R_2^3\rho_2\!-R_1^3\rho_1)/(R_2^3\!-R_1^3)$ 
\begin{equation}
\nonumber
\hat b (\hrho|\rho_{12}\!\lessgtr\!1)\!=\!\frac 1 {\hat \xi}\!\left[\!\frac{\sum_{\rm I,j}{\cal B}(|\rho_{\rm I}-\hrho|\!\leq\!\Delta \rho/2, \rho_{12,\rm I}\!\lessgtr\!1)\rho_{\alpha_{{\rm I},j}}
}{6\sum_{\rm I}{\cal B}(|\rho_{\rm I}-\hrho|\leq \!\Delta \rho/2,\rho_{12,\rm I}\!\lessgtr\!1)}\!-\!1\right]\!, \label{eq:effbrhoconstr}
\end{equation}
 which are both measured with bins of width $\Delta \rho =0.1$.
\subsection{Parametrizing the covariance matrix}
\label{ssec:paramcov}
In order to determine the decay-rate function, the joint PDFs and hence also the bias functions, one needs to compute the covariance matrix between initial densities in spheres of radii $R_{i}$ and $R_{j}$ as defined in equation~\eqref{eq:covmatrix}.
For the sake of simplicity, we choose here to parametrize this covariance matrix in analogy to a power-law initial spectrum with spectral index $n=n(R_p)$ by 
\begin{subequations}
\label{eq:sigijparam}
\begin{align}
\sigma^{2}(R_{i},R_{i})&=\sigma^2(R_{p})\left(\frac{R_{i}}{R_{p}}\right)^{-n(R_{p})-3}\,,\\
\sigma^{2}(R_{i},R_{j> i})
&=\sigma^2(R_{p})\,{\cal G}\left(\frac{R_i}{R_p},\frac{R_j}{R_p},n(R_{p})\right)\,,
\end{align}
\end{subequations}
where
\begin{align}
{\cal G}(x,y,n)&=\frac{\displaystyle \int{\dd^{3}k\,}k^{n}W_{\rm 3D}(k x)W_{\rm 3D}(k y)}{\displaystyle\int{\dd^{3}k\,}k^{n}W_{\rm 3D}(k R_{p})W_{\rm 3D}(k R_{p})}\nonumber
\\
&=
  \! \frac{ (x\!+\!y)^{\alpha} \!\! \left(\!x^2\!+\!y^2\!-\!\alpha x y\right)\!-\!(y\!-\!x)^{\alpha} 
 \!  \!\left(\!x^2\!+\!y^2\!+\!\alpha x y\right)}
   {2^{\alpha}(n+1) x^3 y^3  },\nonumber
\end{align}
with $\alpha=1-n$. The key parameter in the prediction of the PDF is the value of the variance at the pivot scale $R_p$ which we measure in the simulation and use as an input to our theoretical model. We report the results for the measured variance of both the density $\rho$ and the log-density $\mu=\log\rho$ in Table~\ref{tab:variance}.
 
\begin{table}
\begin{tabular}{c||c|c|c|c|c|c|c}
z& $R$ [Mpc/$h$] & 10&11&12&13&14&15 \\\hline\hline
$4.0$ & $\hat\sigma_\rho\simeq\hat \sigma_\mu$ & 0.18 & 0.17 & 0.16 & 0.15 & 0.14 & 0.13 \\\hline
$0.7$ & $\hat\sigma_\rho$ & 0.51 & 0.47 & 0.44 & 0.41 & 0.39 & 0.37 \\
$0.7$ & $\hat \sigma_\mu$ & 0.46 & 0.43 & 0.41 & 0.39 & 0.37 & 0.35\\\hline
$0.0$ & $\hat\sigma_\rho$ & 0.74 & 0.68 & 0.63 & 0.59 & 0.55 & 0.52 \\
$0.0$ & $\hat \sigma_\mu$ & 0.61 & 0.58 & 0.55 & 0.52 & 0.50 & 0.48\\
\end{tabular}
\caption{Variances of the density $\rho$ and the log-density $\mu=\log\rho$ for different radii $R$ and redshifts $z$ as measured from the HR4 simulation.}
\label{tab:variance}
\end{table} 
 
\subsection{The density PDF in concentric spheres}
\label{ssec:PDF-conc}
For the predictions of the one- and two-cell PDF of the density-in-spheres we  use equation~\eqref{eq:saddlePDFN-cell} specialized to an appropriate logarithmic mapping that provides a wide range of applicability for the saddle point approximation. Note that the functional forms were presented in \cite{Uhlemann16} and compared to measurements from a 500 Mpc$/h$ {\tt Gadget2} simulation \citep{gadget2}  sampled with  $1024^3$ particles. Here, we confront them with the significantly more accurate measurements from the HR4 simulation at redshift $z=0.7$; the results for  $z=0$ are shown  in Appendix~\ref{app:figures}.

\subsubsection{One-cell density PDF}

\begin{figure*}
\includegraphics[width=0.95\columnwidth]{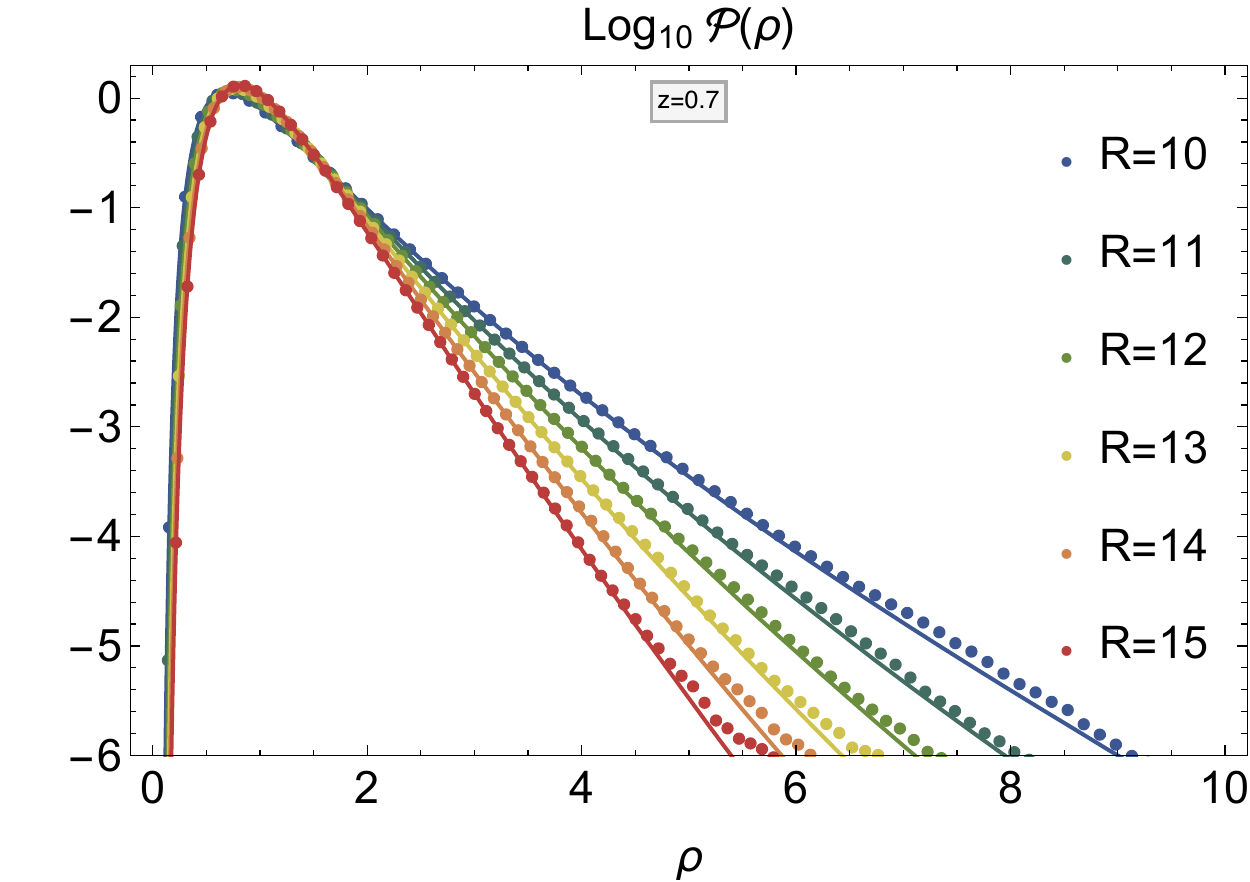}
\includegraphics[width=0.99\columnwidth]{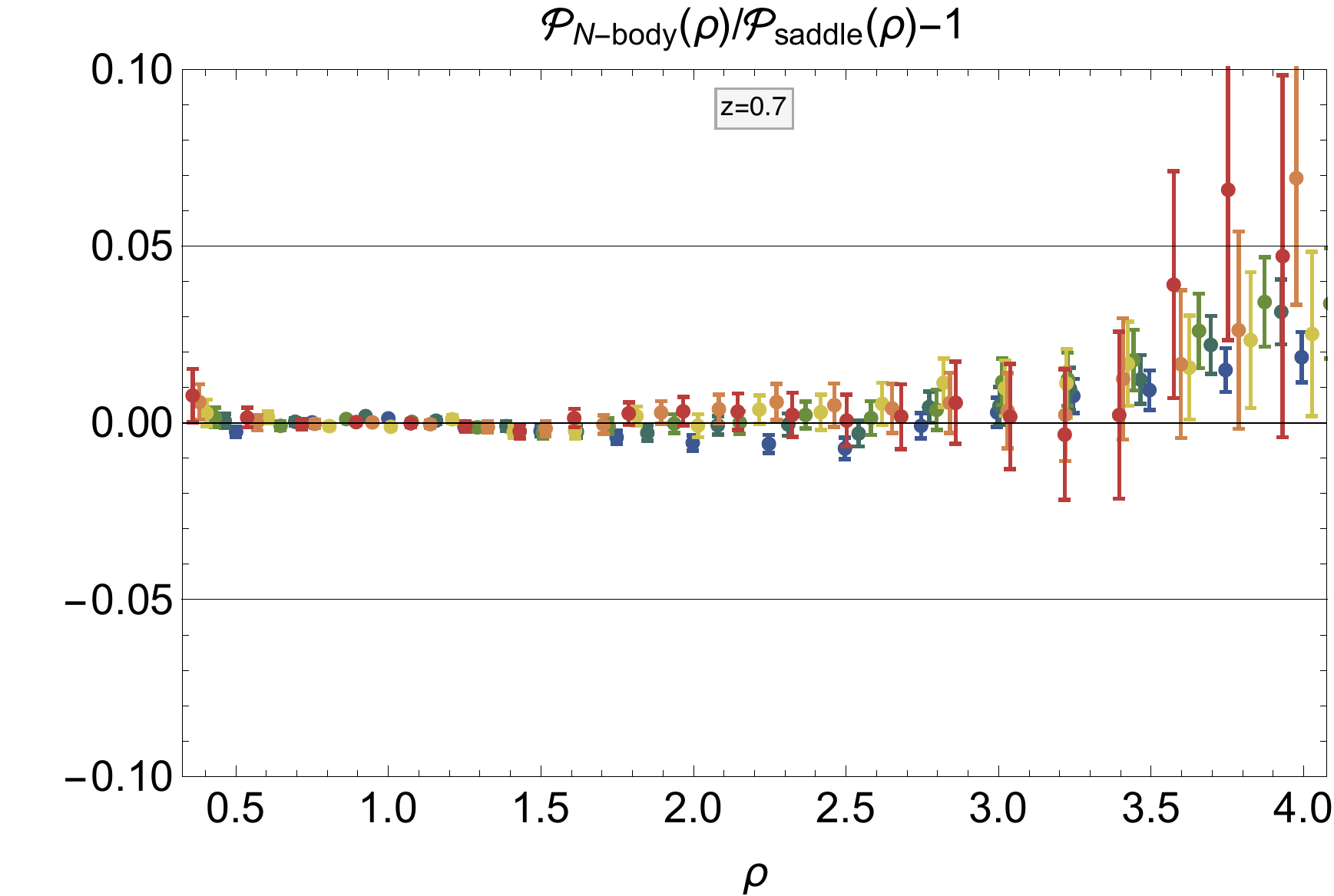}
   \caption{The density PDF $\mP(\rho)$ {\it (left-hand panel)} predicted from the saddle point approximation equation~\eqref{eq:saddlePDF1cell} using {\tt LSSFast} {\it (solid lines)} for redshift $z=0.7$ and radii $R=10,11,...,15$Mpc/$h$ {\it (from blue to red)} with variances as given in Table~\ref{tab:variance}, in comparison to the measurement from HR4 {\it (data points with error bars)} and the corresponding residuals {\it (right-hand panel)}.
     See also Figure~\ref{fig:compPDFsaddleHorizon-z0} for lower  redshift  PDFs.
  {For this figure, we used  $\nu=1.59$ instead of $\nu=21/13\simeq 1.61$ because it leads to smaller residuals (see Appendix~\eqref{app:figures}).}
     } 
   \label{fig:compPDFsaddleHorizon}
\end{figure*}

For the one-cell PDF the appropriate mapping leading to an accurate density PDF is simply the logarithm of the density, 
\begin{align}
\mu = \log\rho \,.
\end{align}
The density PDF is then obtained as
\begin{subequations}
\label{eq:saddlePDF1cell}
\begin{equation}
\label{eq:PDFfromPsi2}
\mP_{R}(\rho) = \sqrt{\frac{\Psi''_R[\rho]+\Psi'_R[\rho]/\rho}{2\pi}} \exp\left(-\Psi_R[\rho]\right)\,.
\end{equation}
The normalized PDF with the corrected mean is obtained from equation\,\eqref{eq:PDFfromPsi2} according to
\begin{align}
\label{eq:PDFfromPsinorm}
\hat\mP_{R}(\rho)= \mP_{R}\left(\rho \cdot \frac{\langle\rho\rangle}{\langle 1\rangle}\right) \cdot \frac{\langle\rho\rangle}{\langle 1\rangle^2}\,.
\end{align}
\end{subequations}

In Figure~\ref{fig:compPDFsaddleHorizon} we compare the saddle point approximations of the PDF obtained from equation~\eqref{eq:saddlePDF1cell} evaluated with the help of {\tt LSSFast} \citep[see also Appendix~\ref{app:LSSfast}]{Codis16b} to the measurements from the HR4 measurement for 6 different radii $R=10,11,...,15$Mpc/$h$ at redshift $z=0.7$.  The agreement is spectacular over a wide range of densities.

\subsubsection{Two-cell joint density slope PDF} 

\begin{figure}
\includegraphics[width=\columnwidth]{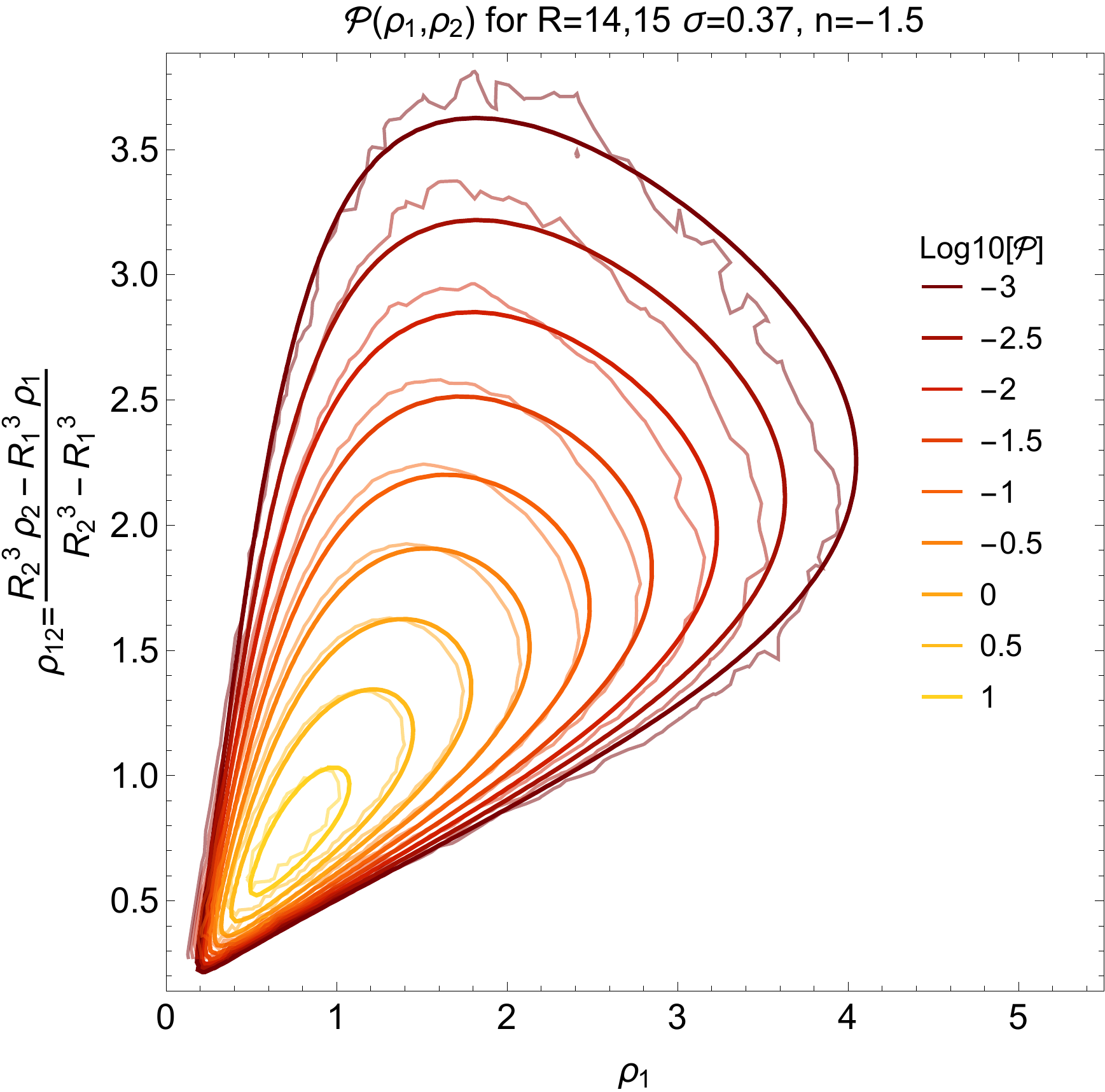}
   \caption{The joint density-slope PDF $\mP(\rho_1,\rho_2)$ predicted from the saddle point approximation equation~\eqref{eq:saddlePDF2cell} {\it (thick lines)} as a function of the central density $\rho_1$ and the shell density $\rho_{12}$ for radii $R_{1,2}=14, 15$ Mpc/$h$ and redshift $z=0.7$ where $\sigma_\mu=0.37$ in comparison to the measurement from HR4  {\it (thin wiggly lines)}. The agreement is quite good and demonstrates the wide dynamical range of this simulation. See also Figure~\ref{fig:compPDFsaddleHorizon-z0} for the same PDF at redshift $z=0$.
  } 
   \label{fig:compjointPDFsaddleHorizon}
\end{figure}

A suitable and physically motivated change of variables for the two-cell case is given by the logarithmic transform of the sum and difference of mass
 \label{eq:eq:log-mass}
 \begin{align}
 \mu_{1}= \log  \left(r^{3}\rho_{2}+\rho_{1}\right)  \,,\quad 
   \mu_{2}= \log \left(r^{3}\rho_{2}-\rho_{1}\right)\,,
 \end{align}
 where the relative shell thickness is $r=R_{2}/R_{1}$ and mass conservation ensures $\mu_{2}$ to be real. The PDF $\mP(\rho_{1},\rho_{2})$ can then be approximated via equation~\eqref{eq:saddlePDFN-cell}, which can explicitly be rewritten as 
 \begin{subequations}
  \label{eq:saddlePDF2cell}
  \begin{equation}
  \label{eq:saddlePDF2cellexpr}
 \mP_{R_1,R_2}(\rho_{1},\rho_{2})=
\frac{ \exp\left[-\Psi_{R_1,R_2}\right]}{ 2 \pi} \sqrt{ p_{R_1,R_2}(\rho_{1},\rho_{2})}\,,
 \end{equation}
 with
 \begin{align}
p_{R_1,R_2}(\rho_{1},\rho_{2})=&\det\left[\frac{\partial^{2}\Psi_{R_1,R_2}}{\partial \mu_{i}\partial \mu_{j}}\right]\left(\det\left[\frac{\partial\mu_{i}}{\partial \rho_{j}}\right]\right)^{2}\\
=
 &
 \left(\frac 1 {2r^{3}} \Psi_{,22}+\Psi_{,12}+\frac {r^{3}} {2}\Psi_{,11}+\frac{\Psi_{,2}+r^{3}\Psi_{,1}}{r^{3}\rho_{2}+\rho_{1}}\right)\nonumber\\
 \times& \left(\frac 1 {2r^{3}} \Psi_{,22}-\Psi_{,12}+\frac {r^{3}} {2}\Psi_{,11}+\frac{\Psi_{,2}-r^{3}\Psi_{,1}}{r^{3}\rho_{2}-\rho_{1}}\right)\nonumber\\
-&\left(\frac {\Psi_{,22}}{2 r^{3}}-\frac{r^{3}\Psi_{,11}}{2}\right)^{2}\nonumber\,,
 \end{align}
  \end{subequations}
with $\Psi_{,1}$ and $\Psi_{,2}$ denoting partial derivatives with regard to $\rho_{1}$ and $\rho_{2}$ respectively. Analogously to the one cell case, one still has to enforce the mean and normalization for the saddle point PDF obtained from equation\,\eqref{eq:saddlePDF2cell} following the procedure described in equations~(\ref{eq:saddlePDFN-cellnorm})-(\ref{eq:saddlePDFN-cellnorm2}). 
In practice, it is often useful to express the joint density PDF $\mP(\rho_1,\rho_2)$ not in terms of the two densities but rather as function of the inner density $\rho=\rho_{1}$ and the slope $s=(\rho_{2}-\rho_{1})/(R_2/R_1-1)$ or the density in the outer shell $\rho_{12}= (R_2^3\rho_2-R_1^3\rho_1)/(R_2^3-R_1^3)$. The result is displayed in Figure~\ref{fig:compjointPDFsaddleHorizon} and compared to the measurements. Overall we observe a very good agreement; it is however not as good as for the one-cell case as we did not compute it with the exact linear power spectrum but assumed a power-law power spectrum through the parametrization \eqref{eq:sigijparam}. It was shown in \cite{Bernardeau14} that taking into account the running of the spectral index can improve the result.

\subsection{The bias functions at large separation}
\label{ssec:biaslarge}

To predict the bias functions for densities-in-spheres, we will use equation \eqref{eq:bsaddle} with the variance of the log-density from Table~\ref{tab:variance} as input together with the normalization procedure described in equation~\eqref{eq:biasnorm}. While the functional form of the density bias was introduced already in \cite{1996A&A...312...11B} for the saddle point approximation applied to the density, we provide improved analytical predictions based on the log-density mapping together with a normalization scheme that are as good as results from a numerical integration presented in \citeauthor{Codis16a} (\citeyear{Codis16a}a).
 Furthermore, we provide new predictions for the joint density bias and the derived constrained biases given the density environment. Here, joint density bias refers to the bias, $b(\rho_{1},\rho_{2})$, corresponding to a region of density $\rho_{1}$ smoothed on $R_{1}$ and $\rho_{2}$ on $R_{2}$.

\subsubsection{One-cell density bias}  
The density bias describes the cross-correlation of spherical cells given one spherically-averaged density at separation $r_e$
\begin{equation}
\label{eq:brho-cross}
1+b(\rho)\xi(r_{e})= \langle\rho'|\rho\, ;r_e\rangle\\
 =\frac{\displaystyle\int_{0}^{\infty}\dd \rho ' \mP(\rho,\rho ';r_{e}) \rho '}
{\mP(\rho)} \,.
\end{equation}
Physically the density bias describes the mean of the density found in a sphere of radius $R$ given that the density in a sphere of the same radius at separation $r_e$ is $ \rho$.

\begin{figure}
\includegraphics[width=1.\columnwidth]{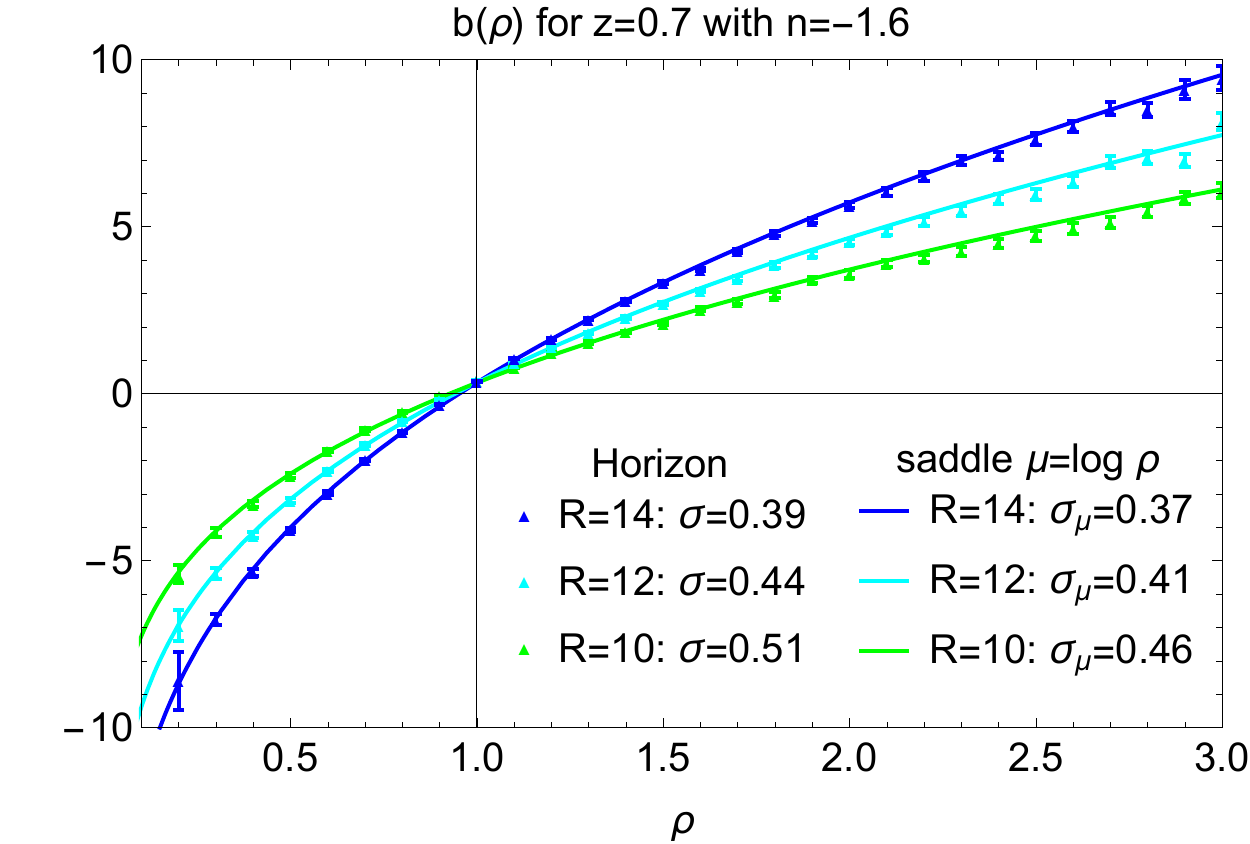}
   \caption{The normalized density bias function $\hat b_R(\rho)$ predicted from the saddle point approximation given by equation~\eqref{eq:densitybias} for the log-density $\mu=\log\rho$ for $n_s=-1.6$ and different values of the variance. This prediction is  compared to the HR4 measurements at redshift $z=0.7$ for different radii $R$ in Mpc/$h$ and hence variances as indicated in the legend. 
   See also Figure~\ref{fig:bias-rho-th-z0} for redshift $z=0$.}
   \label{fig:bias-rho-th}
\end{figure}

\begin{figure}
\centering \includegraphics[width=1\columnwidth]{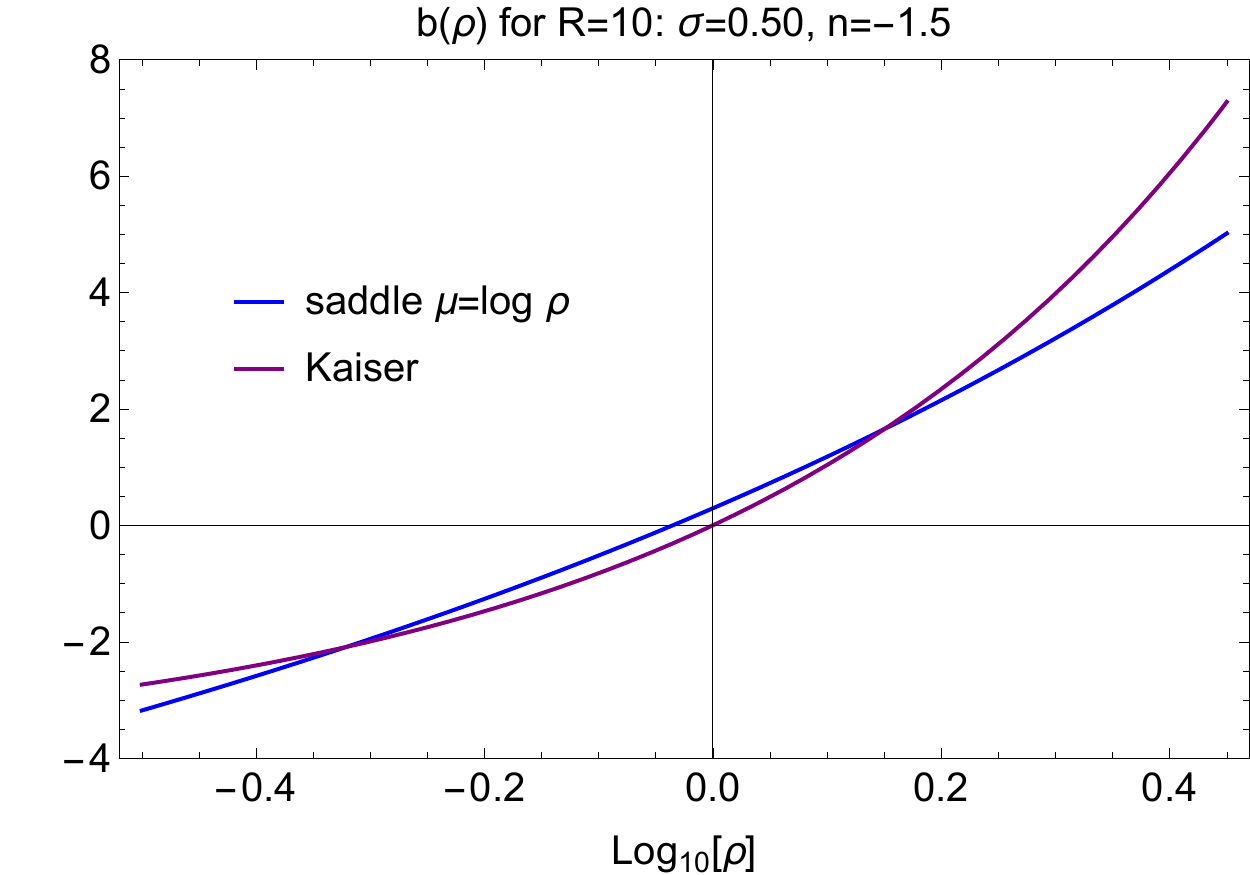}
   \caption{Density bias $b(\rho)$ from equation~\eqref{eq:densitybias} {\it (blue line)} 
   predicted from the saddle approximation for $\sigma=0.50$ and spectral index $n_s=-1.5$ and compared with the Kaiser bias from equation~\eqref{eq:bKaiser} {\it (purple line)} valid only in the Gaussian regime but extrapolated here into the non-Gaussian regime.}
   \label{fig:bias-rho-lim}
\end{figure}

The density bias can be straightforwardly obtained from the saddle point approximation for the log-density $\mu=\log\rho$ by evaluating equation~\eqref{eq:bsaddle} for a given $\rho$ in a cell of radius $R$ as
\begin{subequations}
\label{eq:densitybias}
\begin{align}
\label{eq:densitybias2}
b_R(\rho) &
=  \frac{\nu(1-\rho^{-1/\nu})}{\sigma^2_\mu(R\rho^{1/3})}  \simeq  \frac{\nu(1-\rho^{-1/\nu})\rho^{\frac{3+n(R)}{3}}}{\sigma^2_\mu(R)}  \,.
\end{align}
and then normalizing according to
\begin{align}
\hat b_R(\rho) &= \frac{b_R(\rho)- \langle b_R(\rho)\rangle}{\langle\rho  b_R(\rho)\rangle- \langle b_R(\rho)\rangle}\,. \label{eq:biasnorm1cell}
\end{align}
\end{subequations}
where the averages denoted by $\langle\cdot\rangle$ are computed as integrals with the one-cell saddle point PDF equation~\eqref{eq:saddlePDF1cell}.
The result is plotted in Figure~\ref{fig:bias-rho-th} for redshift $z=0.7$ and radii $R=10,12,14$ Mpc$/h$ with variances as indicated in the legend. In Figure~\ref{fig:bias-rho-slope-th-z0} in Appendix~\ref{app:figures} we also show redshift $z=0$. One can see that unbiased results are obtained for densities close to the mean density, but note that the bias at the  mean density is not exactly zero, as explained in Appendix~\ref{app:biasPT} on the basis of perturbation theory. The amplitude of the bias grows with the deviation from the background density and is, as expected, positive for overdense and negative for underdense regions. {We can clearly see that the saddle point bias functions predict a deviation from the linear growth of Kaiser bias from equation~\eqref{eq:bKaiser1cell} and are in excellent agreement with the measurements;  the normalization procedure \eqref{eq:biasnorm} correctly captures the finite value of the bias at mean density.} In Figure~\ref{fig:bias-rho-lim}, we show that the prediction of equation~\eqref{eq:densitybias}, which relies on large deviation statistics and spherical collapse, extends the classical  Kaiser bias result from equation~\eqref{eq:bKaiser1cell} valid in the Gaussian and hence linear regime towards the non-Gaussian, mildly nonlinear regime. Hence we obtain corrections to the linear growth of the bias with density contrast: we observe that the bias for underdense regions is significantly enhanced because in the non-Gaussian regime the sharp cutoff at zero density becomes apparent which suppresses the rare-event tail further and disfavours very small densities. In turn, the bias for overdense regions is suppressed because the gravitational evolution bends the rare-event tail upwards and therefore favours large overdensities.

\subsubsection{Joint density-slope bias}  

The two-cell bias is given by
\begin{align}
\label{eq:bsrho-cross}
1\!+\!b(\rho_1, \rho_2)\xi(r_{e}) &= \langle\rho'|(\rho_1,\rho_2);r_e\rangle\\
\notag 
&=\frac{\textstyle\int_{0}^{\infty} \dd \rho'\,\mP(\{\rho_1,\rho_2\},\rho';r_{e}) \rho'}{\mP(\rho_1,\rho_2)}\,,
\end{align}
where $\mP(\{\rho_1,\rho_2\},\rho';r_{e})$ is a marginal of the two-cell PDF 
\begin{equation*}
\mP(\{\rho_1,\rho_2\},\rho';r_{e})= \textstyle \int_{0}^{\infty}\dd \rho_2'\, \mP(\{\rho_1,\rho_2\},\{\rho',\rho'_2\};r_{e}) \,.
\end{equation*}
The joint density and slope bias describes the mean of the density found in a sphere of radius $R=R_1$ given that at a distance $r_e$, the densities in spheres of radii $R_{1}$ and $R_2$  are respectively $(\hrho_1,\hrho_2)$.
\begin{figure}
\center
\includegraphics[width=1.\columnwidth]{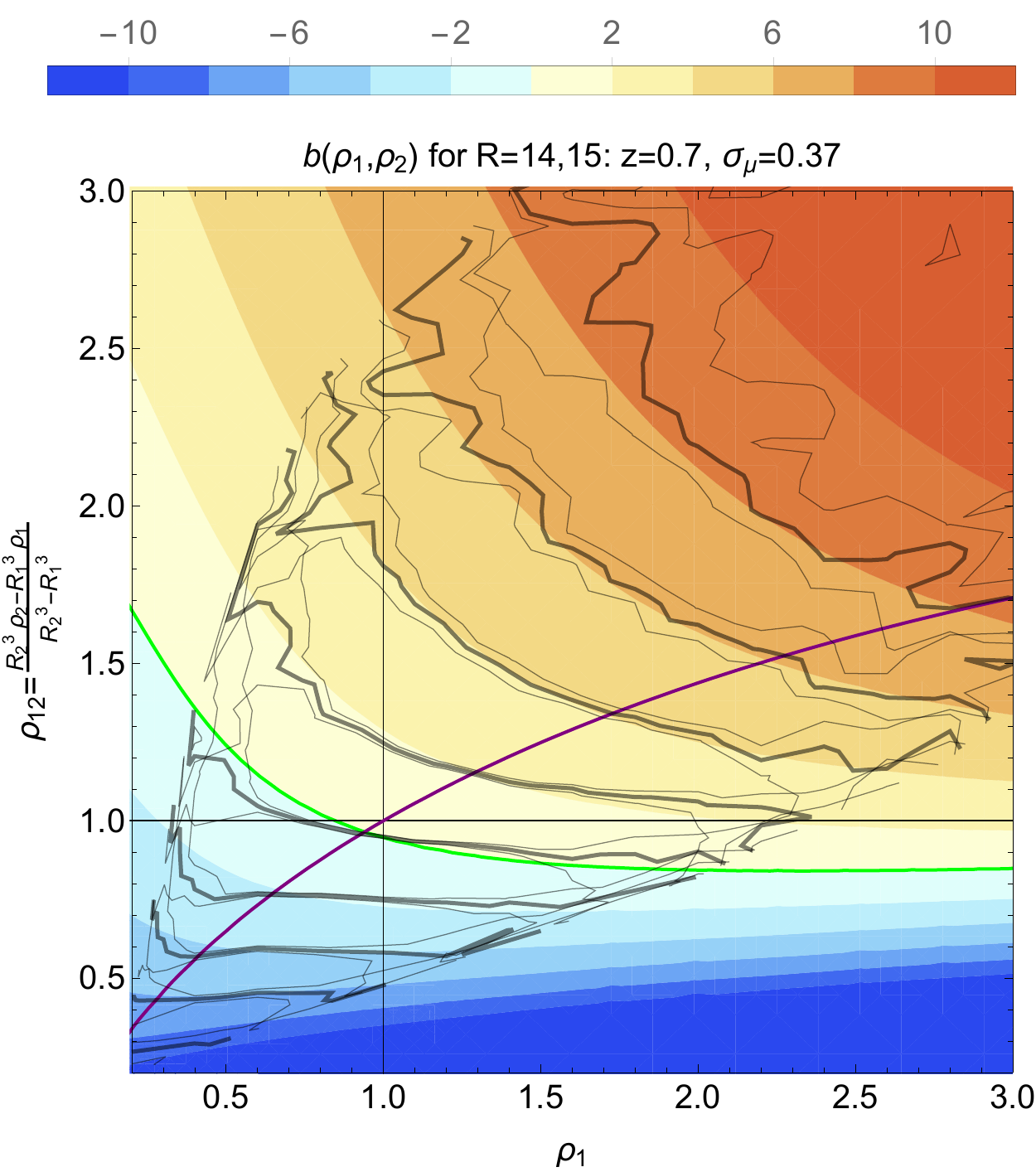}
   \caption{A contour plot of the joint density and slope bias function $b(\rho_1,\rho_2)$ predicted from the saddle approximation equation~\eqref{eq:jointdensityslopebias} with normalization from equation~\eqref{eq:biasnorm} as a function of the central density $\rho_1$ and the shell density $\rho_{12}$ for $R_1=14$Mpc$/h$ and $R_2=15$Mpc$/h$ at redshift $z=0.7$ where $\sigma_\mu$=0.37 (corresponding to $\sigma_\rho=0.39$) in comparison to the measurements from HR4 (mean as thick black lines, and mean $\pm$ error on the mean as thin black lines). Shown is also the stationary line $\rho_2(\rho_1)$ (purple line) along which the joint bias function has to be evaluated to obtain the density bias $b(\rho_1) = b(\rho_1, \rho_2(\rho_1))$. The green line corresponds to zero bias.} 
   \label{fig:bias-rho-slope-th}
\end{figure}
The result, before normalization, is straightforwardly obtained from the general formula~\eqref{eq:bsaddle} 
\begin{align}
\label{eq:jointdensityslopebias}
\hskip -0.25cm b_{R_1,R_2}(\rho_1,\rho_2) &=  \nu \!\!\sum_{i,j=1}^2\! \Xi_{ij}\!\left(R_i\rho_i^{1/3},R_j\rho_j^{1/3}\right) (1-\rho_i^{-1/\nu})\,,
\end{align}
where the covariance matrix is given by equation~\eqref{eq:covmatrix} and can be parametrized by equation~\eqref{eq:sigijparam} for a power-law initial spectrum. We can again use the variables describing the inner density $\rho_1$ together with the slope $s=(\rho_{2}-\rho_{1})/(R_2/R_1-1)$ or the density in the outer shell $\rho_{12}=(R_2^3\rho_2-R_1^3\rho_1)/(R_2^3-R_1^3)$. The result, normalized according to equation~\eqref{eq:biasnorm}, is shown for radii $R_{1,2}=14, 15$Mpc$/h$ at redshift $z=0.7$ in Figure~\ref{fig:bias-rho-slope-th} (and at redshift $z=0$ in Figure~\ref{fig:bias-rho-slope-th-z0}). Besides the general trend that the bias increases with increasing over- or underdensity, one can see that unbiased results are obtained along the green line for which either both densities are close to the background density or the over- or underdensity of the central density is roughly counterbalanced by a under- or overdense shell, respectively. Again we find a deviation from the linear growth predicted by the linear bias described in equation~\eqref{eq:bKaiser} and good agreement with the measurements.

{\it Consistency check.} Note that by decimation of variables we can obtain the one-cell density bias $b(\rho_1)$ from the joint two-cell density bias $b(\rho_1,\rho_2)$ by evaluating it along the stationary line $\rho_{2}^{\text{stat}}(\rho_1)$ of the decay-rate function (shown as purple line in Figure~\ref{fig:bias-rho-slope-th}) as 
\begin{align}
0=\frac{\partial \Psi}{\partial\rho_2}\Big|_{\rho_2=\rho_{2}^{\text{stat}}(\rho_1)} \Rightarrow b(\rho_1)= b(\rho_1,\rho_{2}^{\text{stat}}(\rho_1)) \,,
\end{align}
which indeed gives back the density bias shown in Figure~\ref{fig:bias-rho-th}.

{\it Constrained bias given environment.} Given that we have the full knowledge of the two-cell bias function we can now determine two constrained quantities: the density bias in a positive or negative slope environment $b(\rho_1|\rho_2\gtrless \rho_1)$ as well as the bias in an over- or underdense shell $b(\rho_1| \rho_{12}\gtrless 1)$.

\subsubsection{Density bias in a positive or negative slope environment}

\begin{figure}
\includegraphics[width=0.95\columnwidth]{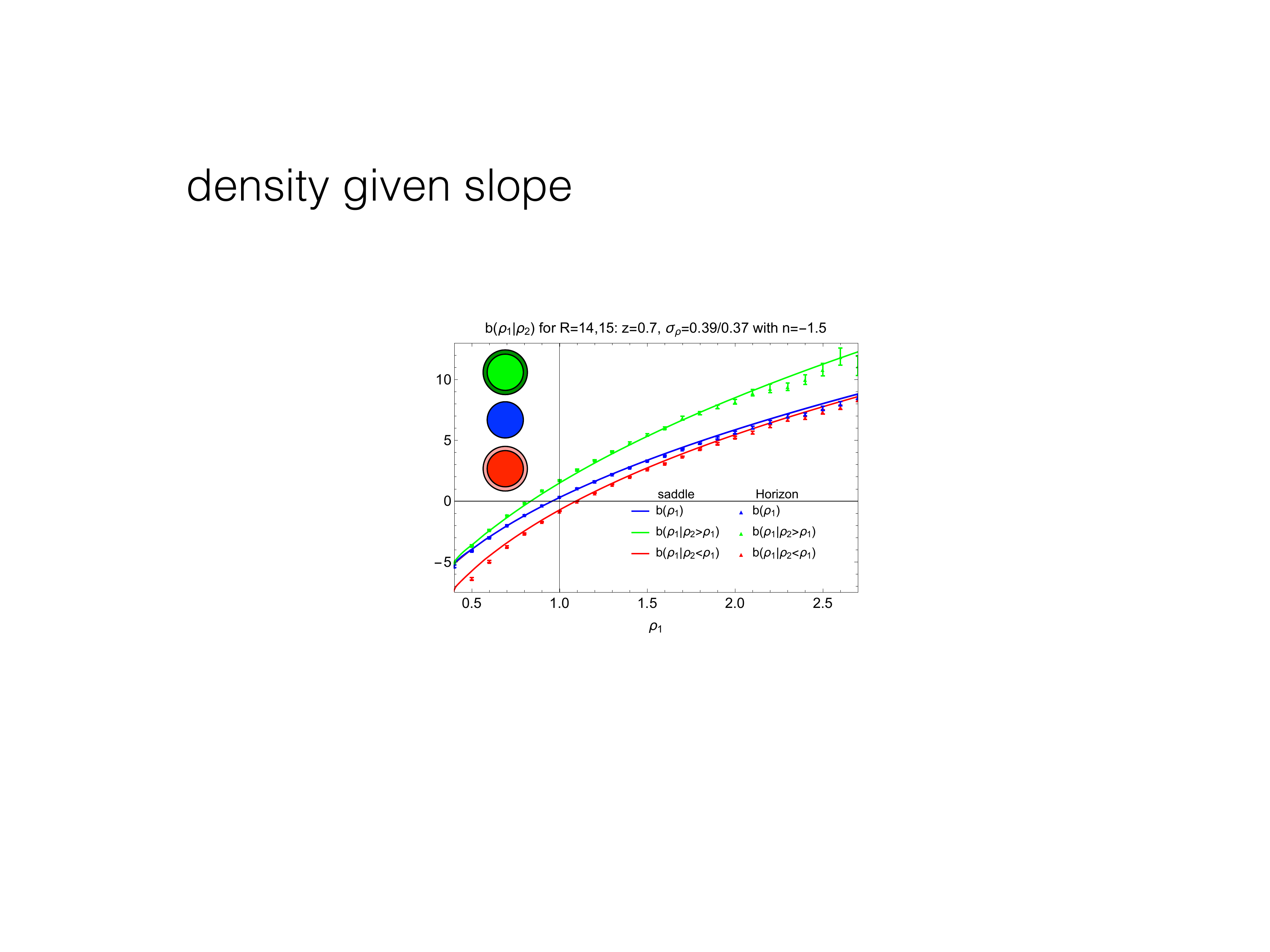}\vspace{0.2cm}\\
\includegraphics[width=0.95\columnwidth]{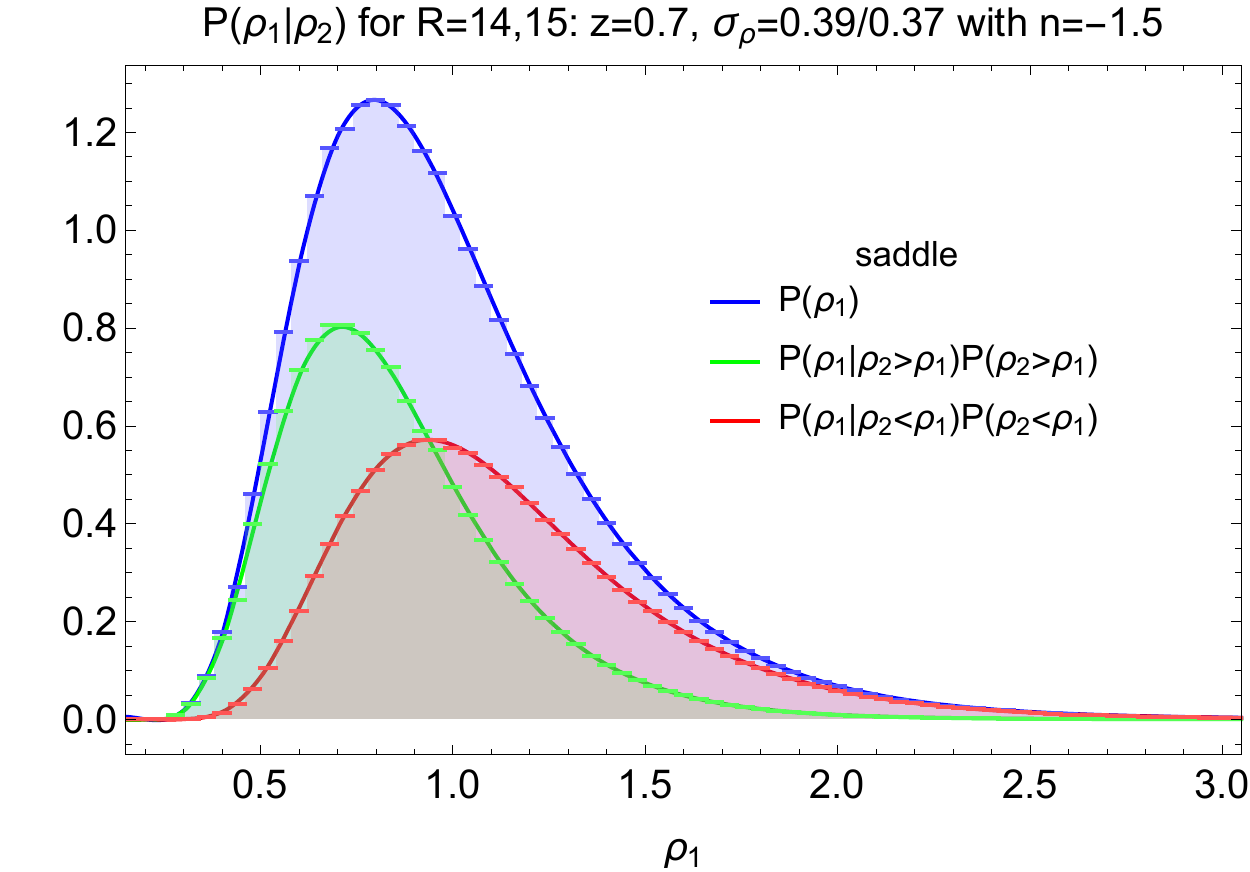}
   \caption{ The constrained density bias function for negative slopes $b(\rho_1|\rho_2<\rho_1)$ {\it (red line)} and positive slopes $b(\rho_1|\rho_2>\rho_1)$ {\it (green line)} compared to the unconstrained density bias $b(\rho)$ {\it (blue line)} which agrees with the density bias from the one-cell saddle point approximation $b_1(\rho)$ {\it (purple line)} shown in Figure~\ref{fig:bias-rho-th}. The coloured spheres, where darker color indicates higher density, sketched in the inset illustrate the different cases considered for the correspondingly coloured lines in the plot. All results are obtained from the saddle point approximation for the log-density for radii $R_1=14$Mpc$/h$ and $R_2=15$Mpc$/h$ at redshift $z=0.7$ with variance $\sigma_\mu$=0.37. 
   Once again the agreement is excellent.} 
   \label{fig:bias-rho-constraint-slope}
\end{figure}

The constrained bias for the density $\rho$ given a positive or negative slope $s$ can be obtained from  
\begin{subequations}
\label{eq:brhoconstrs}
\begin{equation}
(b\cdot\mP)(\rho_1|\rho_2\!\gtrless\!\rho_1) \!=\! \displaystyle \int_{0}^{\infty}\!\!\!\dd \rho_2\; (b\cdot\mP)(\rho_1,\rho_2)\theta(\pm (\rho_2\!-\!\rho_1))\,,
\label{eq:densityconstraint}
\end{equation}
where $\theta$ is the Heaviside step function.
Note that apart from normalization this equation resembles the definition of the constrained PDF when $b=1$ and we have that
\begin{align}
\label{eq:biasrhosplits}
b(\rho_1)\!=\! \frac{(b\cdot\mP)(\rho_1|\rho_2\!<\!\rho_1)}{\mP(\rho_1)}+\frac{(b\cdot\mP)(\rho_1|\rho_2\!>\!\rho_1)}{\mP(\rho_1)}\,,
\end{align}
where we define the constrained bias as
\begin{align}
\label{eq:biasrhosconstraint}
b(\rho_1|\rho_2\!\gtrless\!\rho_1)= \frac{(b\cdot\mP)(\rho_1|\rho_2\!\gtrless\!\rho_1)}{\mP(\rho_1|\rho_2\!\gtrless\!\rho_1)}\,.
\end{align}
\end{subequations}
Hence, we can easily compare the constrained bias function to its unconstrained analogue. This is done in Figure~\ref{fig:bias-rho-constraint-slope}. A positive slope increases the bias for all densities with a strength growing with the central density, while a negative slope has the opposite effect. {Because a mean central density with a negative or positive slope will appear as overall under- or overdense, respectively, the value of the bias at mean density and the point of vanishing bias appear shifted. An intuition about this shift can be gained from a peak-background split argument by computing the mean density given positive or negative slope which yields $\langle\rho_1|\rho_2\!>\!\rho_1\rangle=1.15$ and $\langle\rho_1|\rho_2\!<\!\rho_1\rangle=0.85$, respectively.}
The constrained density bias given slope provides us with a proxy for peaks in the spirit of BBKS, which correspond to overdensities with negative slope (peaks) and underdensities with positive slope (voids), respectively. Those configurations also give the asymptotes of the density bias for extreme densities, because large strongly under- or overdense regions will mostly have positive or negative slopes, respectively, which causes one contribution from equation~\eqref{eq:biasrhosplits} to dominate in the regime of extreme densities. {Note that, while BBKS determine peaks in the initial Gaussian field and then apply collapse criteria based on spherical collapse, we here use spherical collapse to predict the final non-Gaussian statistics of densities-in-spheres and use special configurations to get a proxy to peaks in the final field.} It is worth nothing that  
our formalism describes large peaks of the density field in a similar (but not equivalent) fashion as peak bias studies do following the definition introduced in BBKS  (see e.g \cite{Desjacques10}). Those two different approaches give a complementary insight on halo biasing, either by focusing on overdense regions with negative slopes in the mildly non-linear regime (this work) or on the maxima of the Gaussian Lagrangian density field (BBKS).

\subsubsection{Density bias in an over- or underdense shell}
\begin{figure}
\includegraphics[width=0.95\columnwidth]{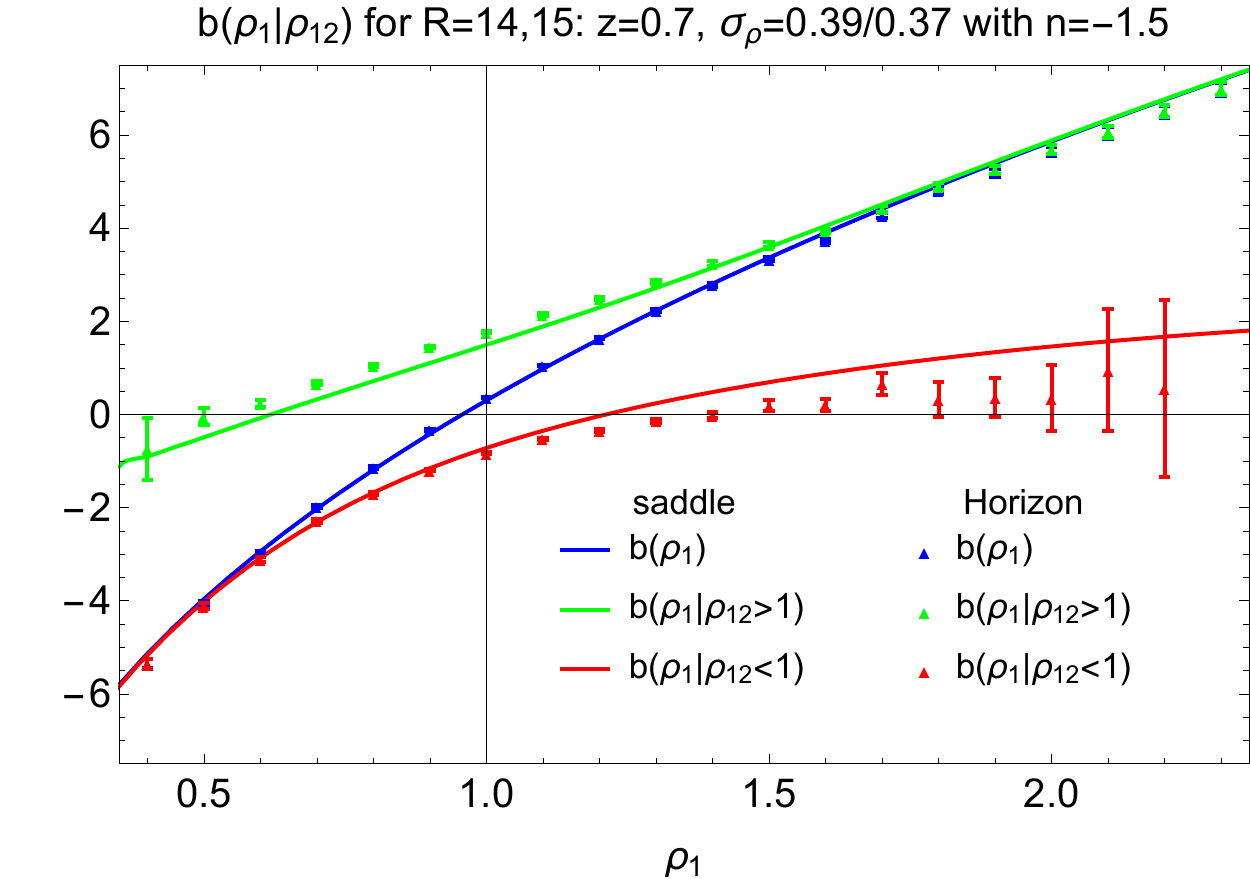}\vspace{0.2cm}\\
\includegraphics[width=0.95\columnwidth]{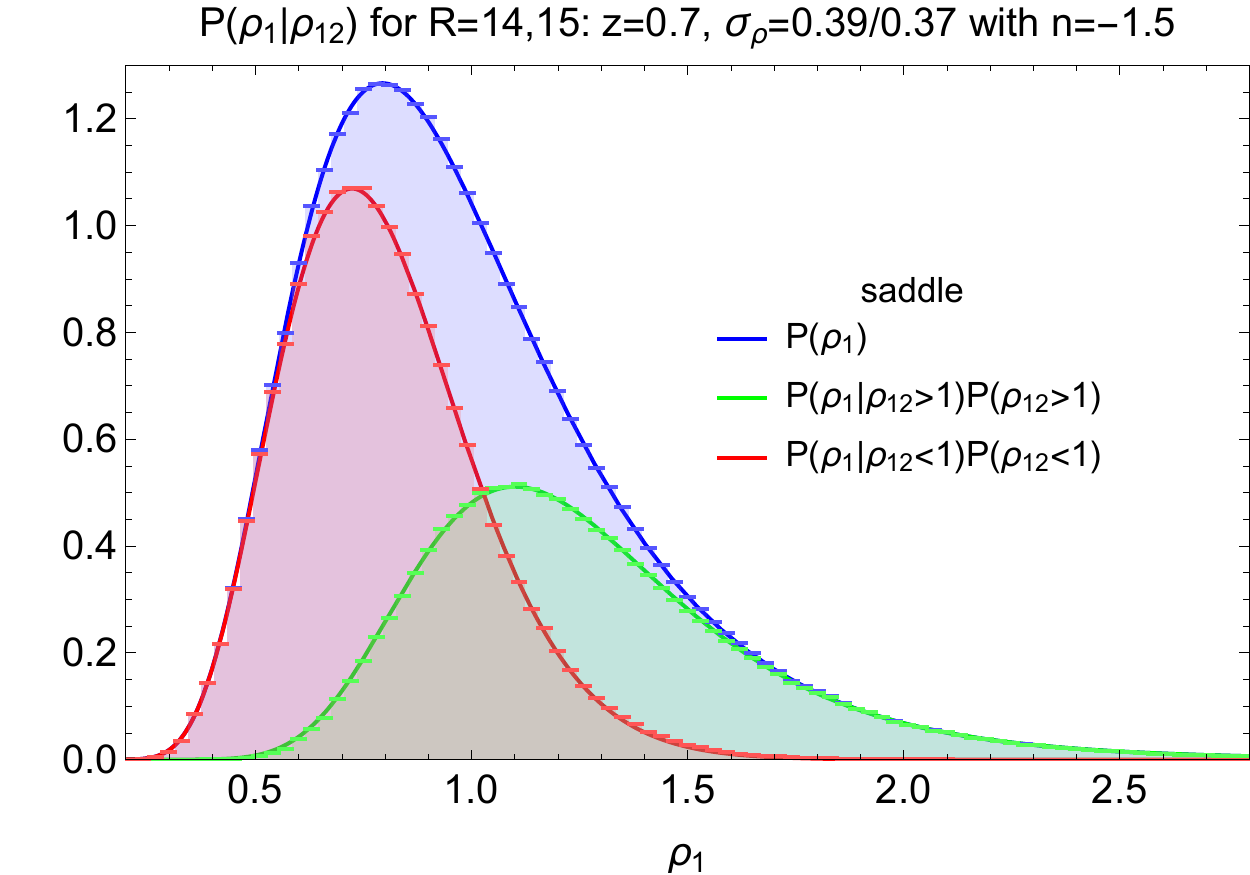}
   \caption{The constrained effective density bias function in underdense shells $b(\rho_1|\rho_{12}<1)$ {\it (red line)} and overdense shells $b(\rho_1|\rho_{12}>1)$ {\it (green line)} compared to the unconstrained density bias $b(\rho_1)$ {\it (blue line)}. All results are obtained from the saddle point approximation for log-density with radii $R_1=14$Mpc$/h$ and $R_2=15$Mpc$/h$ at redshift $z=0.7$ with the variance $\sigma_\mu$=0.37. %
   The agreement is quite good so long as the PDF is significantly non zero.
  } 
   \label{fig:bias-rho1-constraint}
\end{figure}

The constrained bias for the central density $\rho$ given an over- or underdense shell with density $\rho_{12}=(R_2^3\rho_2-R_1^3\rho_1)/(R_2^3-R_1^3)$ can be obtained from  
\begin{subequations}
\label{eq:brhoconstrrho2}
\begin{equation}
\label{eq:densityoverunderdense}
(b\cdot\mP)(\rho_1|\rho_{12}\!\gtrless\!1)\!=\!\!\displaystyle \int_{0}^{\infty}\!\!\!\!\!\dd\rho_{12}\; (b\cdot\mP)(\rho_1,\rho_{12})\theta(\pm (\rho_{12}\!-\!1))\,,
\end{equation}
such that
\begin{align}
\label{eq:biasrhosplitrho2}
b(\rho_1)=\frac{(b\cdot\mP)(\rho_1|\rho_{12}\!<\!1)}{\mP(\rho_1)}+\frac{(b\cdot\mP)(\rho_1|\rho_{12}\!>\!1)}{\mP(\rho_1)}\,.
\end{align}
where we define the effective constrained bias to be
\begin{align}
\label{eq:biasrho21rho12constraint}
b(\rho_1|\rho_{12}\!\gtrless\!1)= \frac{(b\cdot\mP)(\rho_1|\rho_{12}\!\gtrless\!1)}{\mP(\rho_1|\rho_{12}\!\gtrless\!1)} \,.
\end{align}
\end{subequations}
The results are shown in Figure~\ref{fig:bias-rho1-constraint} where we see that the constrained density bias given an underdense- or overdense environment gives the asymptote of the density bias for small and large densities, respectively. This is due to the fact that large, strongly under- or overdense regions will mostly have under- or overdense environment, respectively, which causes one contribution from equation~\eqref{eq:biasrhosplitrho2} to dominate in the regime of extreme densities. The interesting regime is where the constrained biases deviate from the averaged bias which is when we find an overdensity residing inside an underdensity (red line in the half where $\rho_1\!>\!1$)  or vice versa an underdensity residing inside an overdensity (green line in the half where $\rho_1\!<\!1$ ). {In this specific configuration, over- and underdensities are not only peaks, but even isolated, such that we can think of them as voids surrounded by walls or clusters surrounded by voids.} 
{Because a mean central density with a surrounding under- or overdense shell will appear as overall under- or overdense, respectively, the value of the bias at mean density and the point of vanishing bias appear shifted. An intuition about this shift can be gained once again from a peak-background split argument by computing the mean density given an over- or underdense shell which yields $\langle\rho_1|\rho_{12}\!>\!1\rangle=1.3$ and $\langle\rho_1|\rho_{12}\!<\!1\rangle=0.8$, respectively. Besides that, the bias of overdensities is reduced for isolated overdensities because of the `screening' effect of the surrounding underdense shell and vice versa for isolated underdensities.

\subsection{The two-point correlation function of densities-in-spheres}
\label{ssec:corrfct}
The two-point correlation function of densities-in-spheres was introduced in equation~\eqref{eq:jointPDF} to relate the joint PDF of densities at large separation to individual density PDFs. After having obtained analytical predictions for the density bias and constrained density bias we can use them to predict the modulation for the correlation function $\xi(r_e)$ that is introduced by those bias functions. In Figure~\ref{fig:corrfct-spheres} we show the modulation function $b(\rho)b(\rho')=\xi_{\circ}(\rho,\rho';r_{e}) /\xi(r_e)$ computed from the unconstrained density bias and constrained density bias given slope as was shown in Figure~\ref{fig:bias-rho-constraint-slope}. As mentioned before, the constrained density bias given a negative or positive slope can be viewed as giving a proxy to peaks when we have a negative slope around an overdensity (positive peak) or positive slope around an underdensity (negative peak), respectively. Hence, the different bias modulations we show are the ones for the auto- and cross-correlations of masses (unconstrained densities) and peaks (densities with slopes), in the spirit of  \cite{Regos95,Baldauf16}, but with the added value of capturing the quasilinear regime of structure formation.

The upper panel shows the autocorrelation for positive peaks (negative slopes), mass and negative peaks (positive slopes). The mass correlation function in the middle upper panel shows that over- and underdensities among themselves are positively correlated and more strongly clustered than spheres of average density (lower left and upper right part of the plot) while over- and underdense spheres are negatively correlated with each other (lower right and upper left part). {The point of zero bias, when compared to the unconstrained case, is shifted into the quadrant that corresponds to over- or underdensities if positive or negative peaks are involved which also points to the interesting region of the plots where the peak correlation differs from the average correlation.}
The lower panel shows the cross-correlations between positive peaks (negative slopes), mass and negative peaks (positive slopes).
\begin{figure*}
\includegraphics[width=0.65\columnwidth]{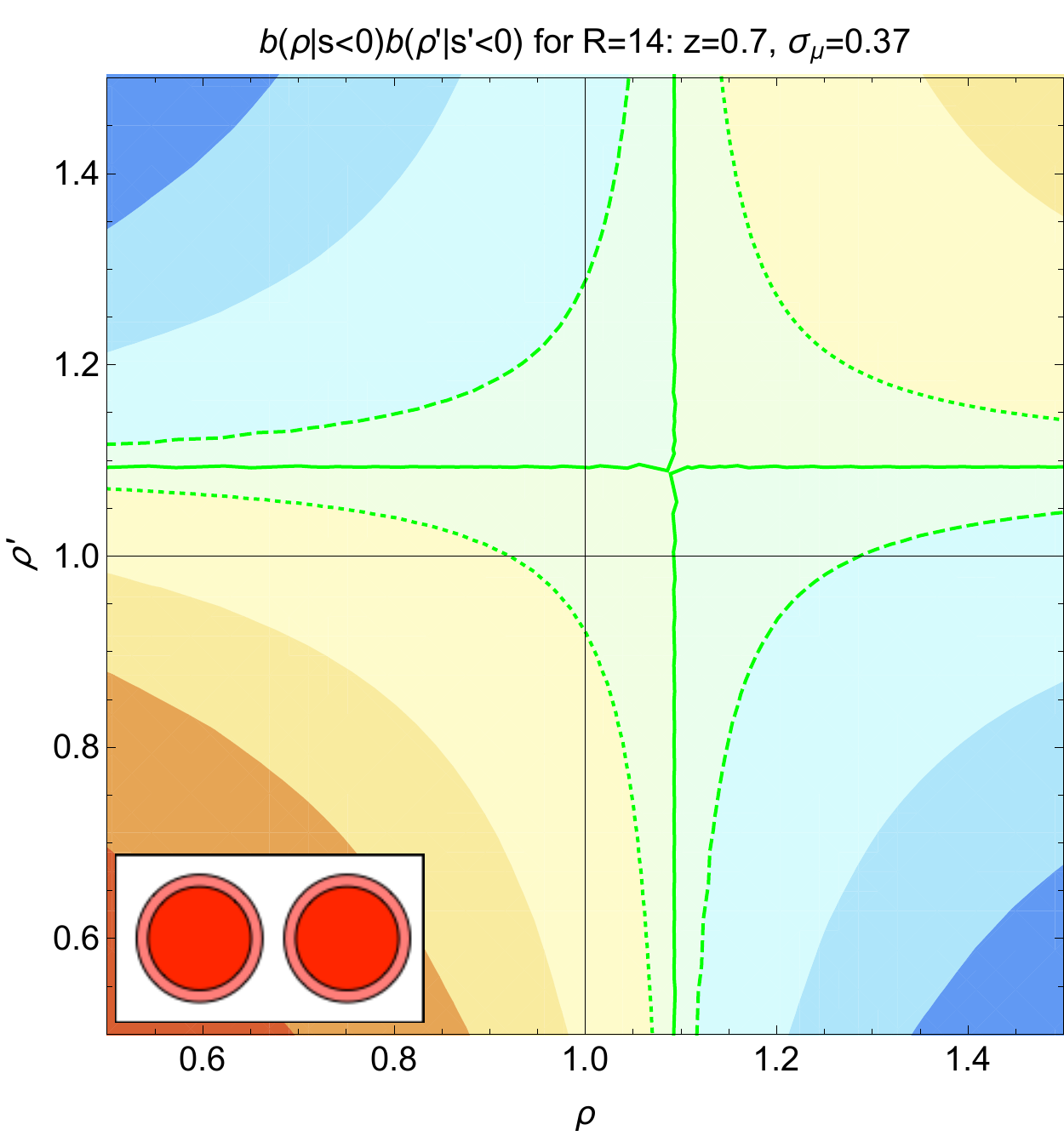}\,
\includegraphics[width=0.65\columnwidth]{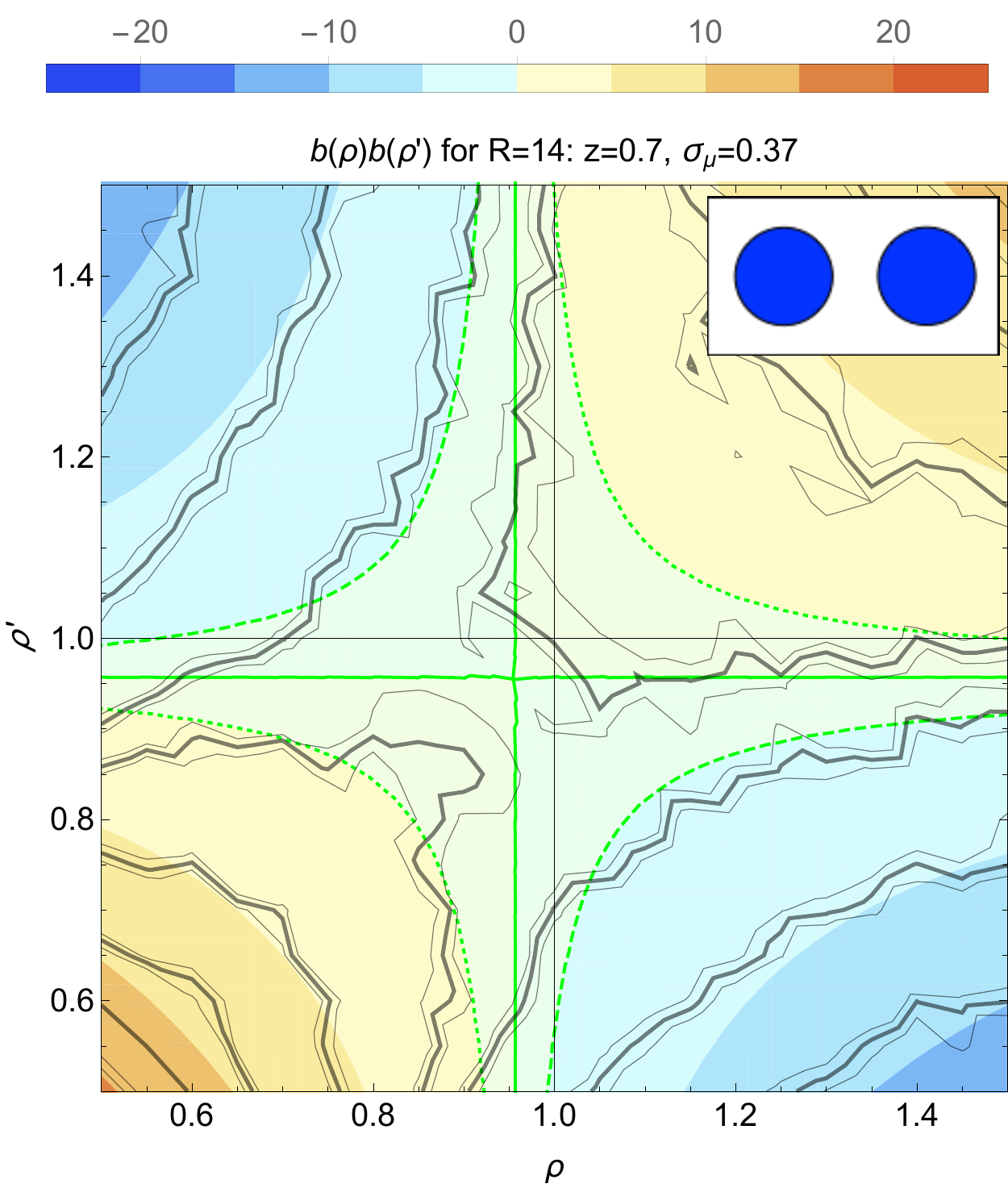}\,
\includegraphics[width=0.65\columnwidth]{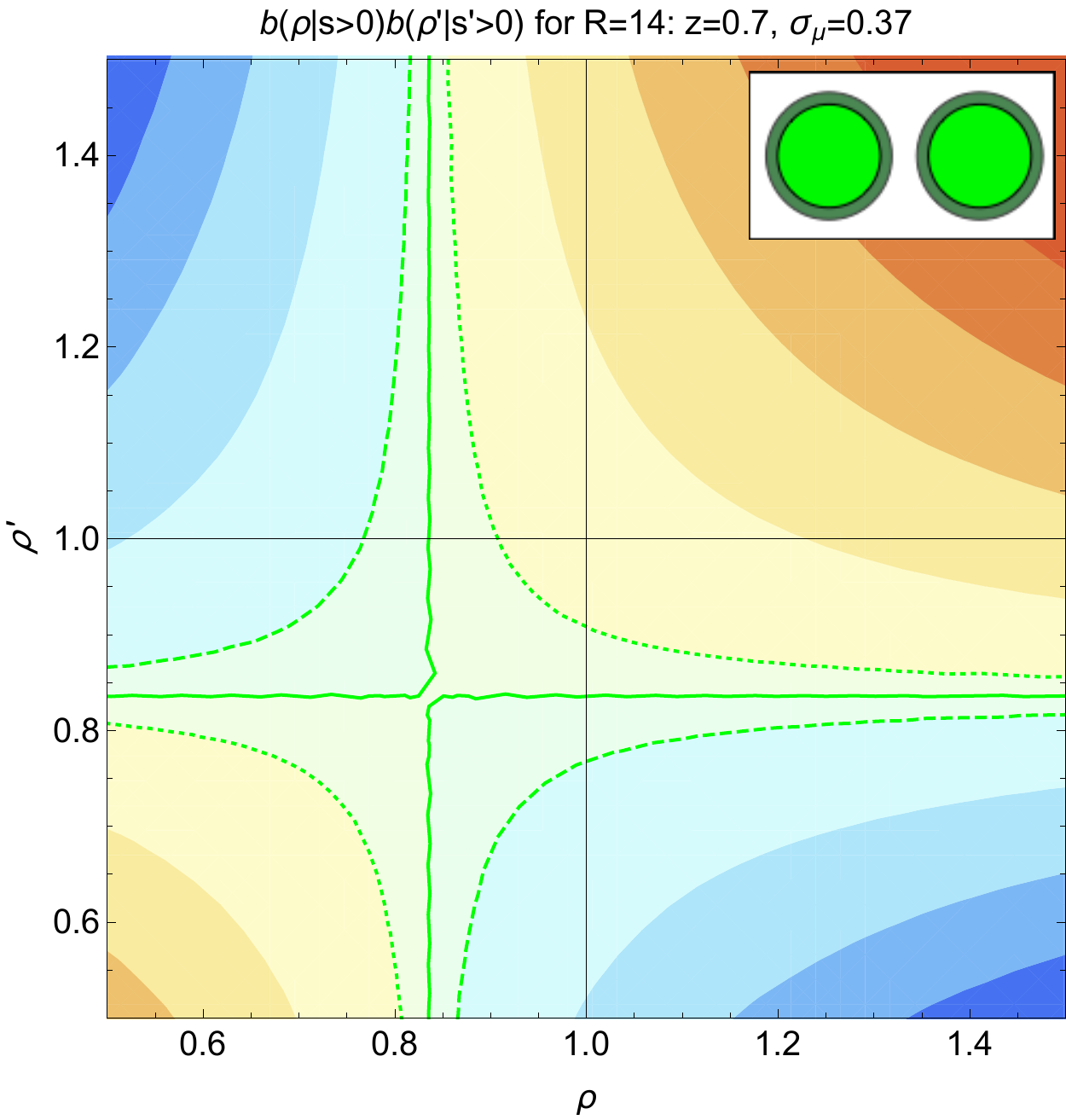}\\
\includegraphics[width=0.65\columnwidth]{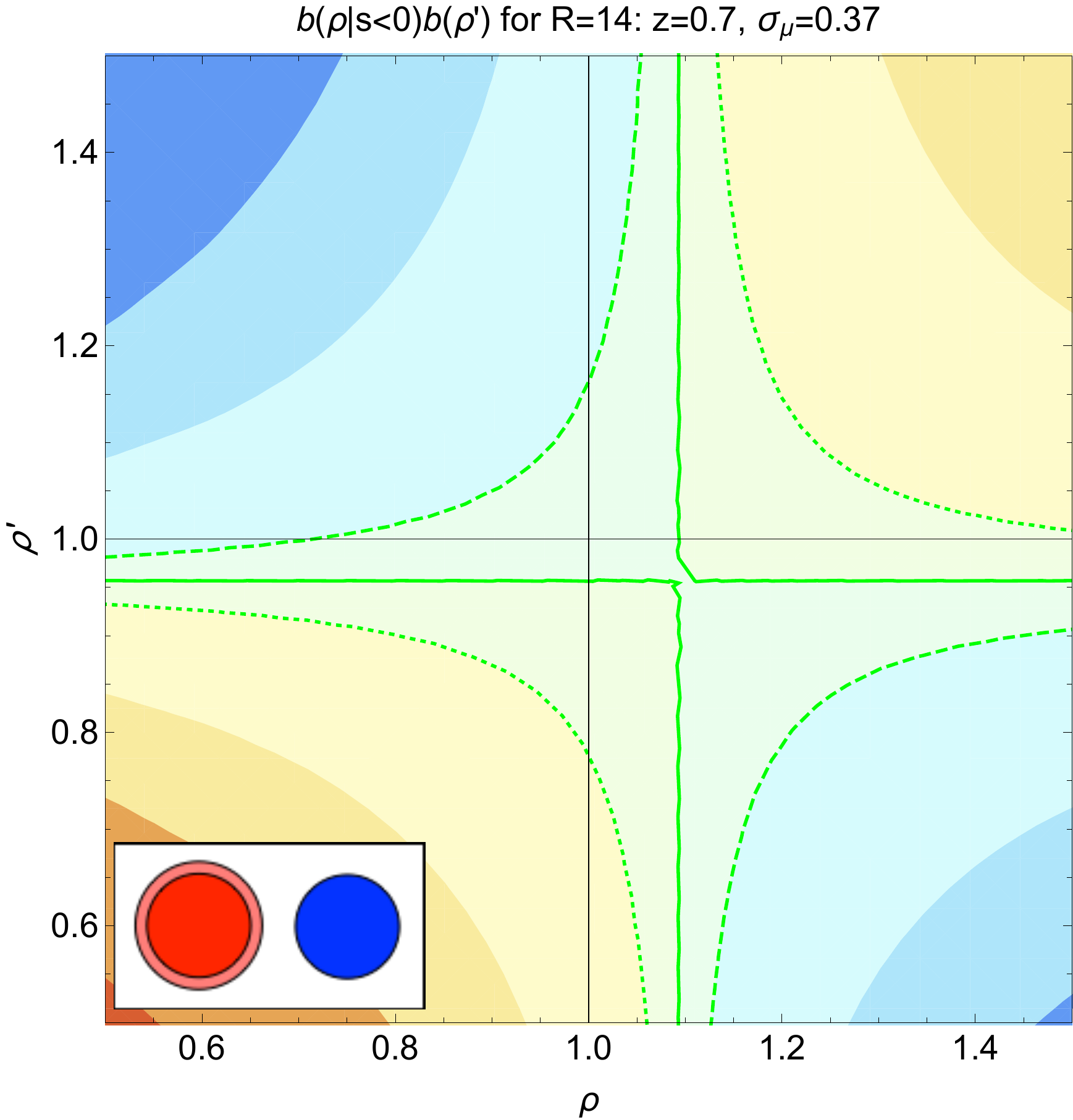}\,
\includegraphics[width=0.65\columnwidth]{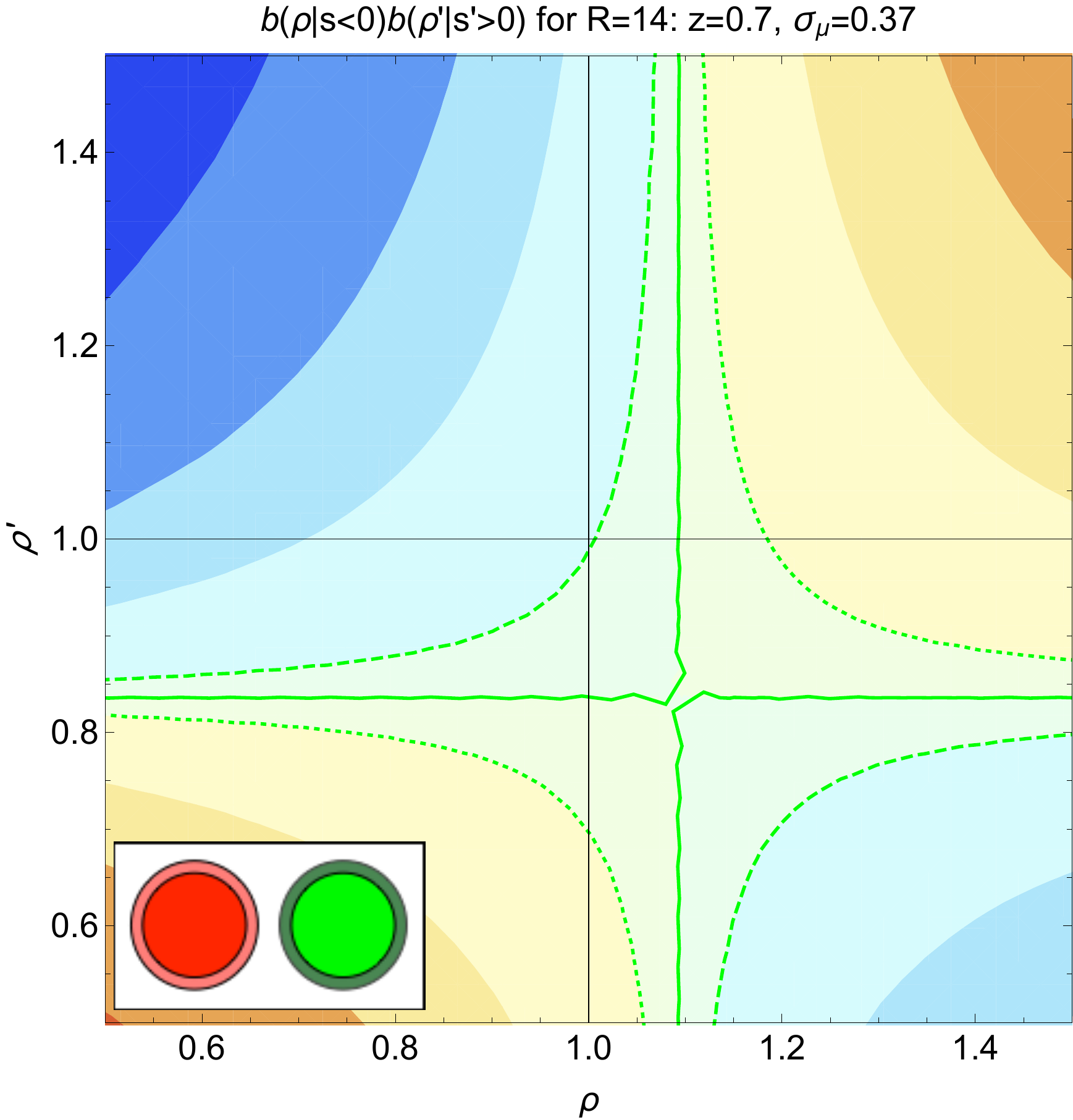}\,
\includegraphics[width=0.65\columnwidth]{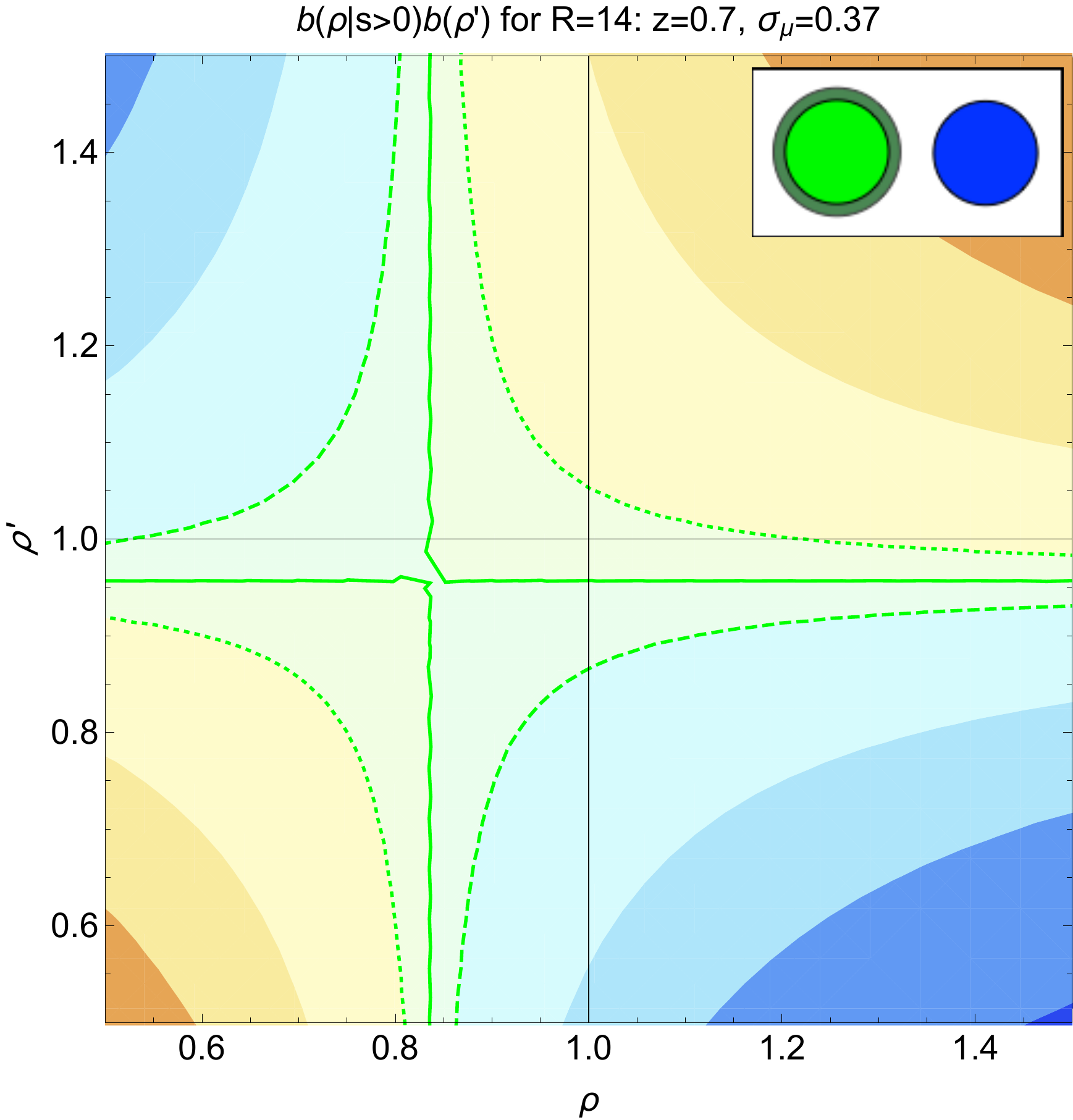}\\
 \caption{The bias modulation of the two-point sphere correlation function $b(\rho)b(\rho')=\xi_\circ(\rho,\rho';r_e)/\xi(r_e)$ for radii  $R=R'=14$Mpc$/h$ at redshift $z=0.7$ with the variance $\sigma_\mu$=0.37 computed from the bias and constrained bias in an over- or underdense mass shell shown in Figure~\ref{fig:bias-rho-constraint-slope}. Besides equi-bias contours we show the line of zero bias $b=0$ {\it (green line)} and unity bias $b=\pm 1$ {\it (green dotted and dashed line)} and sketch the configurations in the insets. 
 {(\it upper panels)} auto-correlations for densities with negative slope (positive peaks) {\it (left)}, unconstrained densities in comparison to HR4 measurements (mean as thick black lines, and mean $\pm$ error on the mean as thin black lines) {\it (middle)} 
 and densities with positive slope (negative peaks) {\it (right)}.
 {(\it lower panels)} cross-correlations between densities with negative slope and unconstrained densities {\it (left)}, densities with negative slope and densities with positive slope {\it (middle)} 
 or densities with positive slope and unconstrained densities {\it (right)}.
 This modulation captures the expected bias clustering of peaks and voids beyond the linear regime.
 }
   \label{fig:corrfct-spheres}
\end{figure*}

Overall the ab initio analytic bias functions for the  two-point correlation of density in spheres, equations~\eqref{eq:densitybias} and \eqref{eq:jointdensityslopebias},
have been shown to be in very good agreement with the HR4 simulation.
This is remarkable, given that these function are very simple explicit algebraic functions of the underlying linear power spectrum 
 via equation~\eqref{eq:covmatrix} or \eqref{eq:sigijparam}.
 It also demonstrates that modern simulations capture very accurately the one and two-point statistics of non-linear gravitational clustering.

\section{Dark matter correlation ML estimator}
\label{sec:recoverxi}

Equation~(\ref{eq:jointPDF0}) allows us to {\it analytically} model the statistics of the cosmic density field in two locations of space.
This model only depends on two parameters: the variance of the density field measured at present-time, $\sigma^{2}(R)$, and the value of the two-point dark matter correlation function, $\xi(r_{e})$, at the separation.
Therefore, following the ideas developed in \citeauthor{Codis16b} (\citeyear{Codis16b}b) for the estimation of the variance $\sigma(R)$, one can build a maximum likelihood estimator for the two-point correlation $\xi(r_{e})$ which should perform better than the sample estimator as time grows and non-gaussianities arise. 

Let us focus here on the two-point density statistics at one scale only
for which  equation~\eqref{eq:jointPDF0} becomes
\begin{equation}
\label{eq:model}
{\cal P}(\rho,\rho')={\cal P}(\rho){\cal P}(\rho')(1+\xi(r_e)b(\rho)b(\rho'))\,,
\end{equation}
where $\rho'$ is the density at a distance $r_{e}$ from $\rho$.
In equation~(\ref{eq:model}), the one-point PDFs only depend on the variance and are computed using the public code {\tt LSSFast} described in Appendix~\ref{app:LSSfast} and the bias $b(\rho)$ is predicted via equation~(\ref{eq:bsaddle}). This two-cell PDF is shown on the left-hand panel of Figure~\ref{fig:estimator} where the effect of the spatial clustering (dashed line) is compared to the case with no spatial correlation (solid line).

 \subsection{Fiducial experiment from HR4}
\label{ssec:fiducial}

Let us carry out the following experiment: consider the $252^{3}$ spheres of radius $R=15$Mpc$/h$ of the HR4 simulation at $z=0.7$ equally spaced on a grid of resolution $\Delta R=12.5$Mpc$/h$ and let us estimate the corresponding 
dark matter correlation function at different separations keeping the variance fixed. To do so, for each separation $r_{i}=4\Delta R\dots 20 \Delta R$, we look for all pairs of spheres separated by $r_{i}$
 and compute the log-likelihood for different models described by $\xi(r_{i})$
 \begin{equation}
{\cal L}(\xi(r_{i}))=\sum_{(p,q)} \log {\cal P}(\rho_{p},\rho'_{q})\,,
\end{equation}
where the indices $(p,q)$ describe all pairs of spheres separated by $r_{i}$.
The maximum of the likelihood can then be found together with the $\alpha$ sigma contours where ${\cal L}=\max {\cal L}(\xi(r_{i}))-1/2\alpha^{2}$.
{In practice, we only consider separations above $4\Delta R=50$Mpc$/h$ to avoid the small region where the modeled two-point PDF is not yet in the large separation regime (but the bias functions,  which describe  the {\sl mean} density at given separation, are in such regime long before that).}
 However, it is expected that one could restrict the analysis around the maximum of the PDF and therefore get an estimate of the dark matter correlation function even for smaller separations. 
 
\begin{figure*}
\includegraphics[width=0.95\columnwidth]{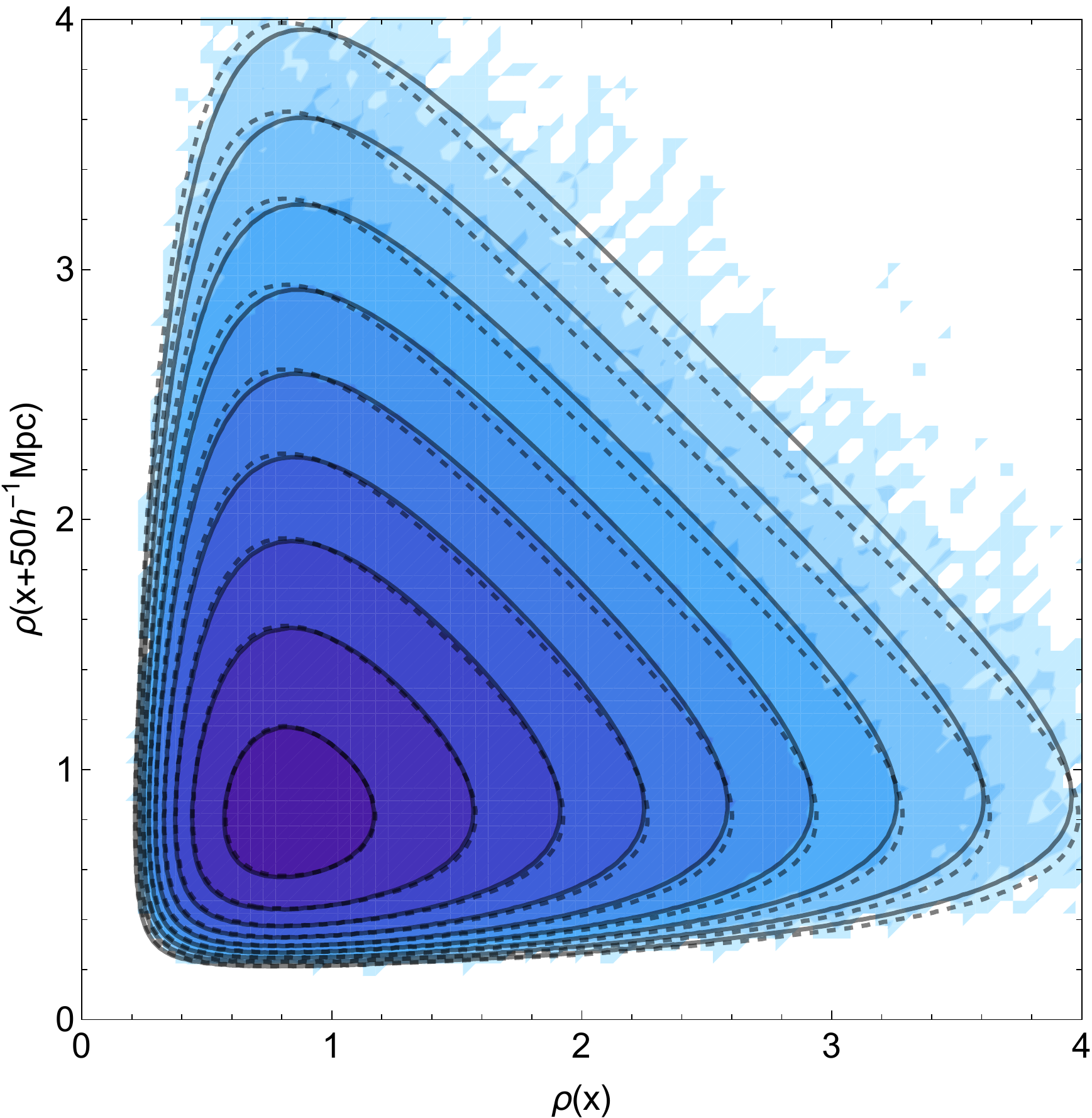}\,
\includegraphics[width=1.02\columnwidth]{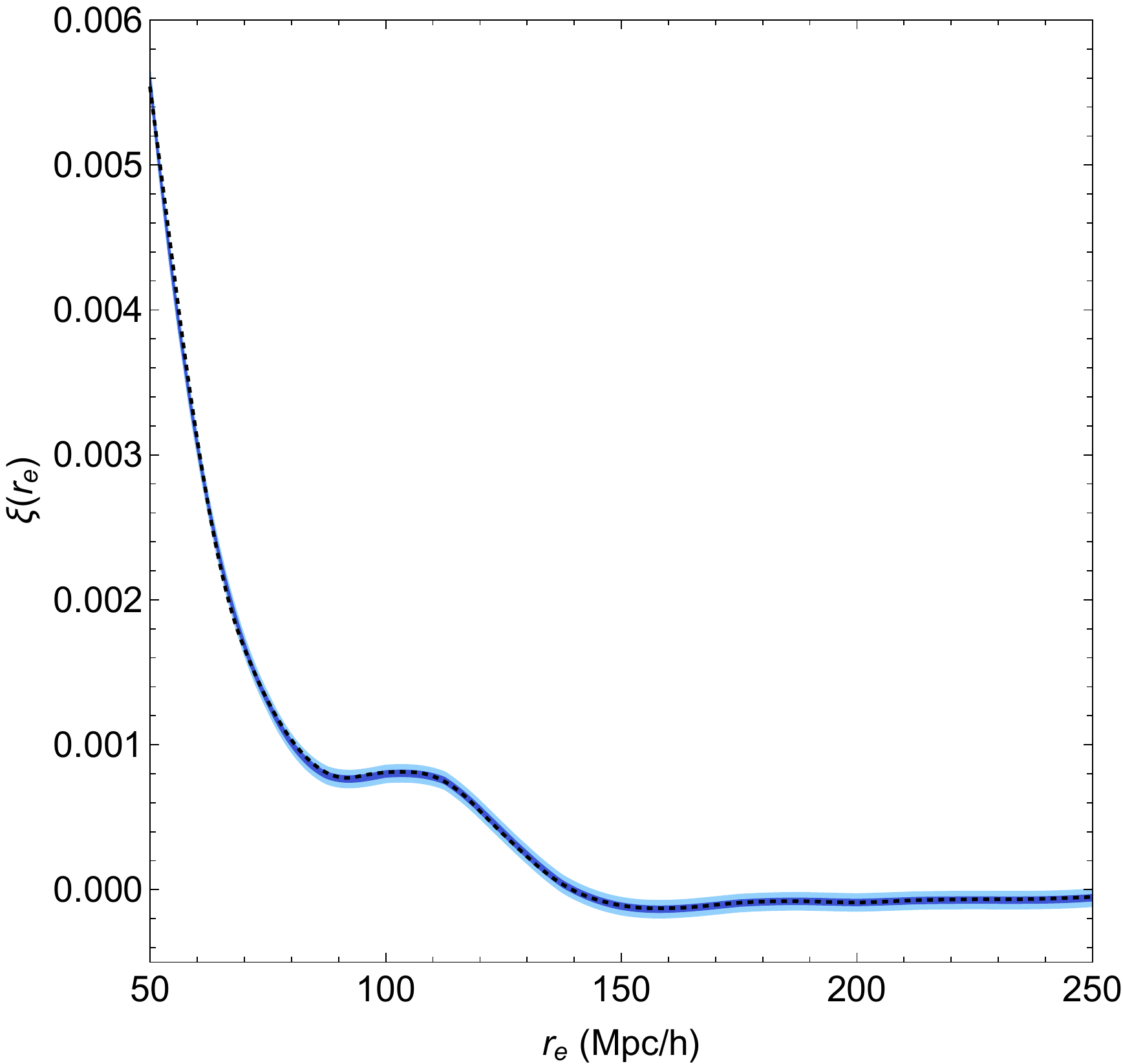}
 \caption{Left-hand panel: PDF of the densities separated by $r_{e}=50$Mpc$/h$ in the HR4 simulation (from dark to light blue) and predicted for $\xi(r_e)=0$ (dashed) and $0.00558$ (dotted). Contours are displayed for ${\log \cal P}=1, 0.5,0 \dots -4$. Right-hand panel: one (dark blue) and three-sigma (light blue) contours
for the maximum likelihood estimate of the dark matter correlation function, $\hat \xi_{\textrm{ML}}$, compared with the arithmetic estimate, $\hat \xi_{\textrm A}$, (dashed line). } 
   \label{fig:estimator}
\end{figure*}

The right-hand panel of Figure~\ref{fig:estimator} shows the corresponding maximum likelihood estimate for $\xi(r_e)$ as a function of the separation compared to a sample estimator with no prior on the underlying PDF
\begin{equation}
\hat \xi_{\textrm A}=\left\langle \rho_{p}\rho_{q}\right\rangle-1\,. 
\end{equation}
The agreement between the sample and likelihood estimators is remarkable, highlighting that the model presented here for the two-point density PDF is very good and could be used to measure more accurately the dark matter correlation function. 

\subsection{Qualifying the estimator}
This procedure is expected to perform better than usual sample estimators when the field becomes mildly non-linear. Indeed, in analogy to the analysis presented in \citeauthor{Codis16b} (\citeyear{Codis16b}b), one can show that the scatter of the maximum likelihood estimator, 
\begin{equation}
\hat\xi_{\textrm{ML}}=\argmax {\cal L}(\xi),
\end{equation}
 is much smaller than that of the sample estimator, $\hat \xi_{\textrm A}$. 
To illustrate this point, for different separations $r_{e}$ between 50 and 250Mpc$/h$, we randomly divide the pairs of spheres of radius $R=15$Mpc$/h$ in 216 subsets. For each subset, we estimate the dark matter correlation function via the sample estimator and the maximum likelihood estimator. The mean and one standard deviation are shown on Figure~\ref{fig:scatter-ML-A}. Both are shown to be unbiased, as the mean is consistent with the correlation function measured from the full simulation, and do not depend on the separation. But, the maximum likelihood estimator is shown to
give a tighter measurement of the dark matter correlation function than the arithmetic estimate, the scatter being reduced by a factor of five in this case. This method could therefore be applied successfully to real surveys provided one is able to model galaxy biasing (Feix et al in prep.).
Such  likelihood estimators could also be generalized  to subset regions of the top-hat filtered field where the density  
has a given value, which could be chosen so as to optimize the  sought level of non-linearity. 

Note finally that the underlying cosmological parameters enter the likelihood function ${\cal L}(z)$ at a given redshift, $z$, twice: via the bias function, $b(\rho,z)$, and via the generalized perturbation theory redshift dependent two-point function, $\xi(r,z)$.
Following \citeauthor{Codis16b} (\citeyear{Codis16b}b), one could imagine in the long run building an optimal dark energy experiment which would measure the 
dark energy parameters while leveraging both the one and two-point statistics 
\begin{equation}
(\hat w_0,\hat w_a)=\argmax_{w_0, w_a} {\sum_z\cal L}(z|w_0,w_a)\,,\notag
\end{equation}
where $w_0$ and  $w_a$ would parametrize the equation of state of dark energy \citep{Glazebrook}.

\begin{figure}
\includegraphics[width=\columnwidth]{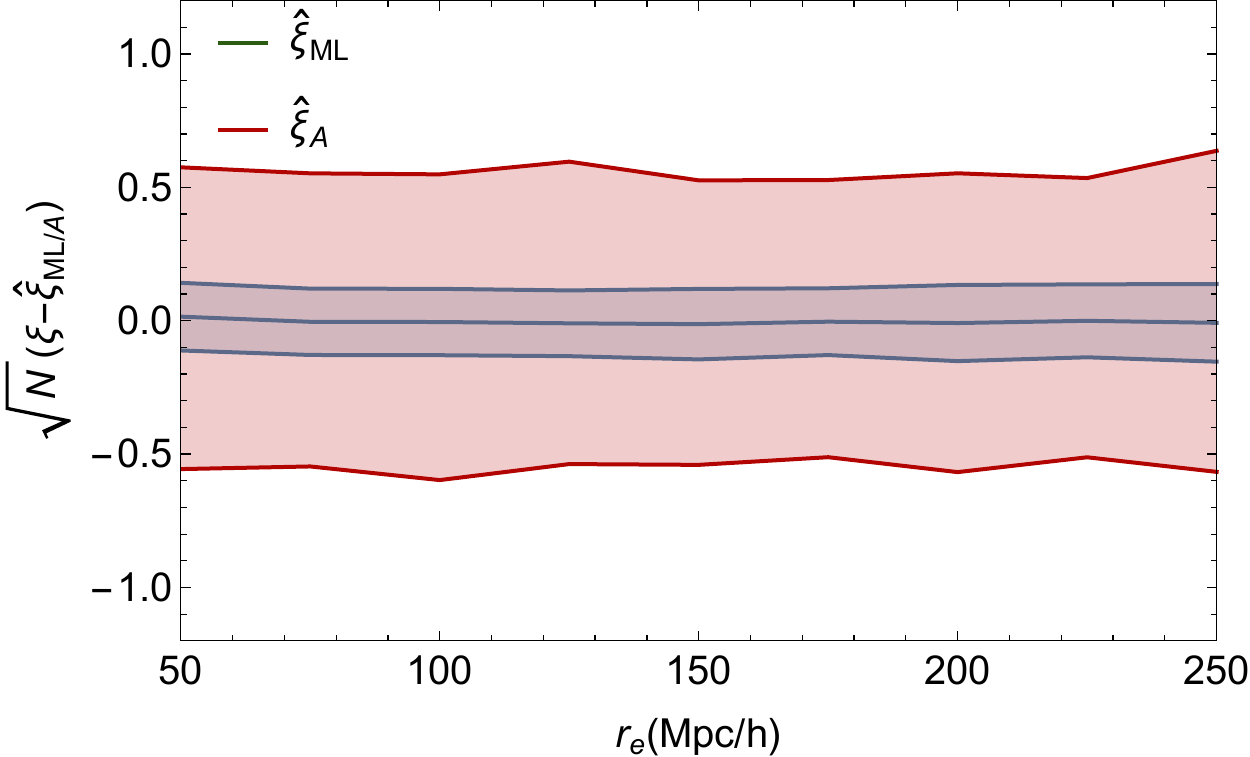}
 \caption{Estimate of the dark matter correlation function at $z=0.7$ in 216 random subsamples of $N=$222,264 pairs of spheres of radius $R=15$Mpc$/h$ separated by $r_{e}=50\dots250$Mpc$/h$. The mean and one standard deviation area are shown in blue for the maximum likelihood estimator  $\hat \xi_{\textrm{ML}}$ and in red for the usual sample estimator $\hat \xi_{\textrm A}$. We display only the difference compared to the dark matter correlation function measured in the full simulation which is assumed to be the true underlying value of $\xi$. As expected, the maximum likelihood estimator provides a much smaller scatter than the sample estimator. } 
   \label{fig:scatter-ML-A}
\end{figure}


\section{Conclusions }
\label{sec:conclusion}

This paper presented 
simple parameter-free analytic bias functions for the  two-point correlation of density in spheres (equations~\eqref{eq:densitybias} and \eqref{eq:jointdensityslopebias}). These bias functions generalize the so-called Kaiser bias in the mildly non-linear regime   for (not so) large separation and  for arbitrary contrast when considering the density smoothed with a top-hat filter (or equivalently measured in spheres). 
 The derivation was carried out using a large-deviation principle, while 
 relying on the spherical collapse model. A logarithmic transformation allowed for a saddle approximation, which was shown to be extremely  accurate against
the state-of-the-art HR4 N-body simulation throughout the range of measured densities, e.g. 
 extending the match to the theory by a factor of 10 or more on joint PDFs, conditionals  and marginals.
 This is both a success of the theory and an assessment of the quality of this simulation.
The conditional density-given-slope and density-given-mass biases were also presented as a quasi-linear proxy to the BBKS extremum correlation functions operating at lower redshifts. As an illustration, Figure~\ref{fig:corrfct-spheres} presented the expected bias modulation of the sphere-sphere correlation function at redshift 0.7 in spheres of 14 Mpc$/h$.

 \citeauthor{Codis16a} (\citeyear{Codis16a}a) recently showed how such bias functions could be used as a mean of mitigating correlation errors when computing  count-in-cell statistics on finite surveys.
 Conversely,   based on the knowledge of the joint PDF of the density in spheres separated by $r_{e}$, we 
 presented and implemented   in Section~\ref{sec:recoverxi} a maximum likelihood estimator for the underlying top-hat smoothed dark matter density, 
 which was shown to be unbiased and very accurate for separations above 50Mpc$/h$.
Its variance  is up to  5 times smaller than that of the classical sample estimator.
Hence these analytic bias functions should be used jointly with analytic models for the two-point function from perturbation theory for cosmic parameter estimation, as they capture the biasing effect of non-linear regime of structure formation. 

Let us stress in closing that the saddle point PDFs presented in this work are not arbitrary fitting functions, but a clear prediction of the theory of gravitational clustering which allows for direct comparison with data a low redshift. These PDFs should be also compared favourably with fits to a lognormal PDF 
 which provide a much worse match as illustrated in Figure~\ref{fig:lognormal-PDF-HR4-z0p7}. The saddle point approximation presented here  gives, at very little extra cost, a few percent accuracy over about 4 orders of magnitude in the values of the one- and two-cell PDFs
and percent accuracy on the bias functions all densities probed by the simulation
 with an explicit dependence of both cosmology, through the initial power spectrum, and the chosen theory 
of gravitation, though the spherical collapse  model. 
In this paper we ignored  redshift-space distortion or galaxy biasing which will be investigated in Feix et al (in prep.).

\vskip 0.5cm
{\bf Acknowledgements:}  
 This work is partially supported by the grants ANR-12-BS05-0002 and  ANR-13-BS05-0005 (\url{http://cosmicorigin.org}) of the French {\sl Agence Nationale de la Recherche}.
 CU is supported by the Delta-ITP consortium, a program of the Netherlands organization for scientific research (NWO) that is funded by the Dutch Ministry of Education, Culture and Science (OCW).
 She thanks IAP for hospitality when this project was completed. 
D. Pogosyan thanks the Institut Lagrange de Paris, a LABEX funded  by the ANR   (under reference ANR-10-LABX-63) within the {\sl Investissements d'Avenir} programme under reference ANR-11-IDEX-0004-02. 
We thank Mark Neyrinck for comments. 

\bibliographystyle{mn2e}
\bibliography{LSStructure}

\appendix

\section{LSSFast Package}
\label{app:LSSfast}
The one-point density PDF and the bias functions for power-law and arbitrary power spectra are made available in the {\tt LSSFast} package distributed freely at \url{http://cita.utoronto.ca/~codis/LSSFast.html}. Two versions of the code are presented. The simpler version, {\tt $\rho$PDFns} and {\tt $\rho$biasns}, assumes a running index, meaning that the variance is given by
\begin{equation}
\sigma^{2}(R)=\frac{2\sigma^{2}(R_{p})}{\left( R /{R_{p}}\right)^{3+n_{s}+\alpha}+\left( R/ {R_{p}}\right)^{3+n_{s}-\alpha}}\,,
\end{equation} 
where $\alpha$ can be non-zero to take into account the variation of the spectral index $n_{s}$. The density PDF is analytically computed from equation~\eqref{eq:saddlePDF1cell} and the bias from equation~\eqref{eq:densitybias}. This code is very efficient and runs in about one second on one processor for one evaluation. Note that the functions {\tt $\rho$PDFns} and {\tt $\rho$biasns} take three arguments, $\rho$, $\sigma$ and $n_{s}$, and has one option, $\alpha$.

The second version of the code, {\tt $\rho$PDF} and {\tt $\rho$bias}, can be applied to arbitrary power spectra. In this case, the function $\sigma^{2}(R)$ is tabulated using equation~(\ref{eq:covmatrix}). Once this tabulation is done (typically one minute on one processor), each evaluation of the PDF and the bias takes about the same time as for the power-law case ($\approx 1$ sec).

\section{PDF and  bias function derivation}
\label{app:LDP}

Let us present shortly the idea behind the large deviation principle that allows to obtain the PDF of densities in concentric cells and the generalization to the joint PDF of densities at different positions. {For more details we refer to \cite{Uhlemann16} and  \citeauthor{Codis16a} (\citeyear{Codis16a}a)}.

\subsection{The PDF of density in concentric spheres}
 \cite{Bernardeau14} computed the joint PDF, $\mP(\{\rho_{k}\})$, of densities in concentric spheres, a highly symmetric configuration which allows to take advantage of the spherical collapse model for gravitational dynamics. To obtain the PDF we use the cumulant generating function of densities in concentric cells, $\varphi(\{\lambda_{k}\})$, defined via a Laplace transform of the density PDF $\mP(\{\rho_k\})$
\begin{align}
\label{phidef}
\varphi(\{\lambda_{k}\})& = \log\left[\int \Pi_k \dd\rho_k \exp(\Sigma_k \lambda_k\rho_k) \, \mP(\{\rho_k\})\right]\,,\\
\notag &= \log [\langle\exp({\Sigma_k} \lambda_k\rho_k)\rangle] =\sum_{p_{i}=0}^{\infty}\ \langle \Pi_{i}\ {\rho_{i}}^{p_{i}} \rangle_{c}\frac{\Pi_{i}\lambda_{i}^{p_{i}}}{\Pi_{i}p_{i}!}\,.
\end{align}
This relationship is useful because, in the limit of zero variance, the cumulant generating function is obtained analytically from the decay-rate function $\Psi(\{\rho_{k}\})$ via a Legendre transformation
\begin{equation}
\varphi(\{\lambda_{k}\})=\sum_{i}\lambda_{i}\rho_{i}-\Psi(\{\rho_{k}\})\,, \, \lambda_{i}=\frac{\partial}{\partial\rho_{i}}\Psi(\{\rho_{k}\})\,,\label{eq:phi2psi}
\end{equation}
where the conjugate variables $\{\lambda_{k}\}$ are functions of the densities $\{\rho_{k}\}$ via  the stationary condition on the decay-rate function which in turn has been obtained from the initial decay-rate function by a simple remapping according to spherical collapse as described in equation~\eqref{PsiDef} (as a result of the contraction principle).
The PDF of the density is then given as an inverse Laplace transform of the cumulant generating function  $\varphi(\{\lambda_{k}\})$
\begin{equation}
 \label{eq:PDF}
\mP(\{\rho_{k}\})=
\int \prod_k \frac{\dd \lambda_k}{2\pi \ii} \exp\left[{ -}\sum_k \lambda_{k}\rho_{k}{ +}\varphi(\{\lambda_{k}\})\right]\,.
\end{equation}
Hence, PDF can be obtained from a numerical integration in the complex plane as done in \cite{Bernardeau14,Bernardeau15} or evaluated using a saddle point approximation for the log-density (which has a close-to-optimal range of validity) as described in \cite{Uhlemann16}.

\subsection{Two-point clustering of concentric spheres}
Let us now consider two sets of concentric spheres separated by a distance $r_e$ and define the corresponding densities $\{\rho_{k}\}\equiv\{\rho_{1,k}\}$ and $\{\rho'_{k}\}\equiv\{\rho_{2,k}\}$. We are interested in the joint density PDF $\mP(\{\rho_{k}\},\{\rho_{k}'\};r_e)$ which, at large separations $r_e>R_k$, can be predicted from the individual PDFs $\mP(\{\rho_{1/2,k}\})$ and some effective bias functions $b(\{\rho_{1/2,k}\})$ according to equation~\eqref{eq:jointPDF}. The effective bias functions encode the correlations of the densities in spheres which are hence related to the two-point correlation function $\xi$ of the underlying dark matter distribution. In analogy to the density PDF at one point, also the joint density PDF at two different points can be obtained from the corresponding cumulant generating function $\varphi(\{\lambda_{k}\},\{\lambda_{k}'\};r_{e})$ of the joint cumulants $\langle 
\rho_{1}^{p_{1}}\dots\rho_{n}^{p_{n}}
{\rho'}_{1}^{q_{1}}\dots{\rho'}_{m}^{q_{m}}
\rangle_{c}$ as inverse Laplace transform
\begin{align}
 \label{eq:PDF}
&\mP(\{\rho_{k}\},\{\rho_{k}'\};r_e) =\\
\notag &\int \prod_k \!\frac{\dd \lambda_{k}}{2\pi \ii}\!  \frac{\dd \lambda_{k}'}{2\pi \ii}\!  \exp\left[-\Sigma_k(\lambda_{k}\rho_{k} +\lambda_{k}'\rho_{k}') +\varphi(\{\lambda_{k}\},\{\lambda_{k}'\};r_e)\right]\,.
\end{align}
In \citeauthor{Codis16a} (\citeyear{Codis16a}a) it has been shown that in the large separation limit, where the separation
distance $r_e$ is much larger than all radii $R_k$ at the individual points, the joint cumulant generating function $\varphi(\{\lambda_{k}\},\{\lambda_{k}'\};r_e)$ can be derived based on the following idea: for large enough separations the joint cumulants can be shown to be well approximated by
\begin{align}
&\langle 
\rho_{1}^{p_{1}}\dots\rho_{n}^{p_{n}}
{\rho'}_{1}^{q_{1}}\dots{\rho'}_{m}^{q_{m}}
\rangle_{c}=
\nonumber\\&\hspace{1cm}
\frac{1}{\xi(r_{e})}
\langle 
\rho_{1}^{p_{1}}\dots\rho_{n}^{p_{n}}
{\rho'}_{1}
\rangle_{c}
\langle 
\rho_{1} {\rho'}_{1}^{q_{1}}\dots{\rho'}_{m}^{q_{m}}
\rangle_{c}\,. \label{eq:cummixed}
\end{align}
Hence, in this limit, it is enough to know the subset of cumulants of the type $\langle \rho_{1}^{p_{1}}\dots\rho_{n}^{p_{n}} {\rho'}_{1} \rangle_{c}$ to determine the generating function of joint cumulants $\varphi(\{\lambda_{k}\},\{\mu_{k}\};r_e)$. The generating function of the joint cumulants of this special type can be shown to be
\begin{equation}
\label{eq:phib}
\varphi_{b}(\{\lambda_{k}\},r_{e})=1+\xi(r_{e}) b_{\varphi}(\{\lambda_{k}\})\,,
\end{equation}
with the bias cumulant generating function defined as
\begin{equation}
b_{\varphi}(\{\lambda_{k}\})\equiv\sum_{i=1}^{n} \sum_{j=1}^{n}\Xi_{ij}\tau_{j}\,.  \label{eq:defbphi}
\end{equation}
Equations~(\ref{eq:cummixed}) and (\ref{eq:phib}) can then be used to express 
the joint cumulant generating function $\varphi(\{\lambda_{k}\},\{\lambda_{k}'\};r_e)$ in terms of  the already known generating function $\varphi(\{\lambda_{k}\})$ of cumulants at one point via
\begin{align}
&\varphi(\{\lambda_{k}\},\{\lambda_{k}'\};r_{e})= 
\nonumber\\
&\hspace{.4cm}\varphi(\{\lambda_{k}\}){+} \varphi(\{\lambda_{k}'\}) +
\xi(r_{e})\, b_{\varphi}(\{\lambda_{k}\}) \, b_{\varphi}(\{\lambda_{k}'\}) \,. \label{eq:phirhokrhokp}
\end{align}
The bias cumulant generating function, $b_{\varphi}(\{\lambda_{k}\})$, is therefore defined as the sum of the first partial derivatives of the initial decay-rate function and hence closely related to equation~(\ref{PsiDef}).  The bias function is then obtained from the bias cumulant generating function via 
\begin{align}
b(\{&\rho_{k}\})\, \mP(\{\rho_{k}\})=\nonumber\\
&\int \prod_k \frac{\dd \lambda_{k}}{2\pi \ii}  b_{\varphi}(\{\lambda_{k}\})
\exp\left(\displaystyle { -}\sum_k \lambda_{k}\rho_{k}{ +}\varphi(\{\lambda_{k}\})\right) \,.  \label{eq:defbiasPDF}
\end{align}
Evaluating the integral in equation~\eqref{eq:defbiasPDF} using a saddle point approximation then gives
\begin{align}
\label{eq:blim}
b(\{\rho_{k}\}) 
&\approx b_{\varphi}\Bigg(\Bigg\{\lambda_{k}=\frac{\partial \Psi(\{\rho_{i}\})}{\partial\rho_k} \Bigg\}\Bigg)  = \sum_{i,j=1}^n \Xi_{ij}\tau_i(\rho_i)\,.
\end{align}

\section{Weakly non-Gaussian bias}
\label{app:biasPT}
Let us revisit here the origin of the normalization shift discussed in the main text by looking at the Gaussian and weakly non Gaussian 
predictions for the mean density bias.
\subsection{Kaiser bias}
To study one-cell and two-cell bias, we first
diagonalise the covariance matrix from equation~\eqref{eq:Kaisercovmatrix},
by transforming from $\tau_1^\prime, \tau_1,\tau_2$ set of correlated 
variable to the following set of independent variables
\begin{align}
\label{eq:decorr}
\nu_{1}&=\frac{\tau_{1}}{\sigma_{11}}\,,\quad \zeta= \frac{ \sigma_{11}^2 \tau'_{1}-\xi_{11}\tau_{1}}{\sigma_{11} \sqrt{ \sigma_{11}^{4}-\xi_{11}^{2}}}\,,\\
\eta&=\frac{\sigma_{11}}{\sqrt{\sigma_{22}^{2}\sigma_{11}^{2}-\sigma_{12}^{4}-\alpha^{2}}}\left(\tau_{2}-\frac{\sigma_{12}^{2}}{\sigma_{11}^{2}}\tau_{1}-\frac{\alpha}{\sigma_{11}}\zeta\right)\,,
\end{align}
which are built to be decorrelated and normalised by their variance: 
\begin{align*}
\left\langle \nu_{1}^{2}\right\rangle=\left\langle \zeta^{2}\right\rangle=\left\langle \eta^{2}\right\rangle=1\,,\quad
\left\langle \nu_{1}\zeta\right\rangle=\left\langle \nu_{1}\eta\right\rangle=\left\langle \zeta\eta\right\rangle=0
\end{align*}
once $\alpha$ is set to $\alpha=(\xi_{12}\sigma_{11}^{2}-\xi_{11}\sigma_{12}^{2})/\sqrt{\sigma_{11}^{4}-\xi_{11}^{2}}$.
Thanks to the diagonalization $(\nu_{1},\zeta,\eta)$ now follow a standard normal distribution, such that it is easy to check that the density bias reads
\begin{align}
b^{\rm G}(\tau)=\frac{\left\langle \tau'_{1}(\zeta,\nu_{1})|\nu_{1}=\tau/\sigma_{11}\right\rangle}{\xi_{11}}
=\frac{\tau}{\sigma_{11}^{2}}\,,
\label{eq:biasG}
\end{align}
which is proportional to the initial overdensity $\tau$ as expected from \citep{Kaiser84}. The two-cell density bias also follows as
\begin{align}
b^{\rm G}(\tau_{1},\tau_{2})&=\frac{\left\langle \tau'_{1}(\zeta,\nu_{1})|(\tau_{1},\tau_{2})\right\rangle}{\xi_{11}}  \approx \sum_{i,j=1}^2 \Xi_{ij}\tau_j\,,
\end{align}
if we assume that for large separations $r_{e}$ we have $\xi_{12}\approx \xi_{11}$.

\subsection{Expected offset of density bias}
Figure~\ref{fig:bias-rho-th}
shows that for the non-linear density field, the one cell
bias $b(\rho)$ is non-vanishing and positive
at the mean density $\rho=1$ whereas the Gaussian result from
equation~(\ref{eq:biasG}) predicts zero bias.
Here we compute this offset using perturbative 
methods in the weakly non-Gaussian regime which are expected to be accurate
for variances of order $\sigma \lesssim 0.2 $.

We use a 
moment expansion for the two-cell distribution function.
The Gaussian limit provides the kernel to define the
orthogonal polynomials of the expansion \citep{Gay2012}.
In the decorrelated variables $\nu_1$ and $\zeta$ introduced in equation~\eqref{eq:decorr} these are just
products of Hermite polynomials, which for the first non-Gaussian
correction to the two-point distribution function give
\begin{align}
P(\nu_{1},\zeta) &\approx \frac{1}{2\pi} \exp\left(-\frac{\nu_1^2}{2} - \frac{\zeta^2}{2}\right) \\
&\times \Big[ 1  + \frac{1}{6} \left\langle \nu_1^3 \right\rangle H_3(\nu_1) 
+ \frac{1}{6} \left\langle \zeta^3 \right\rangle H_3(\zeta) \nonumber\\
&\quad + \frac{1}{2} \left\langle \nu_1 \zeta^2 \right\rangle H_1(\nu_1) H_2(\zeta) 
+ \frac{1}{2} \left\langle \nu_1^2 \zeta \right\rangle H_2(\nu_1) H_1(\zeta) 
\Big]. \nonumber
\end{align} 
The conditional mean that determines the one-cell bias is
\begin{equation}
\langle\tau_1^\prime | \tau_1=\tau \rangle = 
\frac{ \displaystyle \int_{-\infty}^\infty \!\!\!\!\!\! d\nu_1
\int_{-\infty}^\infty \!\!\!\!\!\! d\zeta
\, \tau_1^\prime(\nu_1,\zeta) 
P(\nu_1,\zeta) \delta_D(\tau_1(\nu_1,\zeta) - \tau) }
{\displaystyle \int_{-\infty}^\infty \!\!\! d\nu_1
\int_{-\infty}^\infty \!\!\! d\zeta \; P(\nu_1,\zeta) 
\delta_D(\tau_1(\nu_1,\zeta) - \tau) } \nonumber
\end{equation}
where inverting equation~\eqref{eq:decorr} gives
\begin{equation}
\tau_1 = \nu_1 \sigma_{11} ~, \quad
\tau_1^\prime = \frac{ \nu_1 \xi_{11} +  \zeta \sqrt{\sigma_{11}^4 - \xi_{11}^2}}
{\sigma_{11}}\,.
\end{equation}
After some algebra,
and expressing the moments of 
$\nu_1, \zeta$ back via the moments of $\tau_1, \tau_1^\prime$, 
the one-cell bias at $\rho = \bar\rho=1$ in the leading non-Gaussian order
becomes
\begin{equation}
b(\rho=1) \sim \frac{1}{2}\left( \frac{\langle\tau_1^3\rangle}{\sigma_{11}^4} 
- \frac{\langle\tau_1^\prime \tau_1^2\rangle}{\sigma_{11}^2\xi_{11}(r)}
\right) \label{eq:defbias6}
\end{equation}
As expected, it is zero for $r=0$. In the large separation limit, 
the cubic moments can be computed using perturbation
theory {\citep{Bernardeau02}} and give
\begin{equation*}
\frac{\langle \tau_1^3 \rangle}{\sigma_{11}^4} = S_3 \approx \frac{34}{7} - (n+3)
~, \quad 
\frac{\langle \tau_1^\prime \tau_1^2\rangle}{\sigma_{11}^2 \xi_{11}}
= C_{21} \approx \frac{68}{21} - \frac{1}{3} (n+3)
\end{equation*}
such that the bias offset evaluates to 
\begin{equation}
b(1) \approx \frac{1}{2} ( S_3 - C_{21} )=  -\frac{4}{21} -\frac{n}{3} \stackrel{n=-1.6}{\approx} 0.34\,,
\end{equation}
a value fully consistent with the measured one.

\section{Redshift zero match}
\label{app:figures}

While we focused our comparison between the theoretical predictions and the HR4 simulation in Section~\ref{sec:validation} to redshift $z=0.7$, we here provide results from high redshift $z=4$ until today $z=0$ to outline the reach of our formalism. 

\subsection{Saddle point vs. lognormal PDF}
The one-cell saddle point PDF presented in equation~\eqref{eq:saddlePDF1cell} obtained from a log-density mapping has to be contrasted to an {ad-hoc} lognormal PDF
\begin{equation}
\mP_{\text{lognorm}}(\rho,\sigma) =\frac{1}{\sqrt{2\pi\sigma}} \frac{1}{\rho} \exp\left[-\frac{(\log\rho +  \sigma^2/2)^2}{2 \sigma^2}\right]
\end{equation} 
with a best fit for the variance $\sigma$ which mis-matches the simulated PDFs at the 10\% level or more in its tail, as shown in Figure~\ref{fig:lognormal-PDF-HR4-z0p7}.  This is to be compared with the excellent match seen in Figure~\ref{fig:compPDFsaddleHorizon}. Note that, while doing a joint fit of the mean (which is otherwise assumed to be $\sigma^2/2$) and variance does improve the fit around the mean density, it worsens the mismatch in the tail.

\begin{figure}
\includegraphics[width=1\columnwidth]{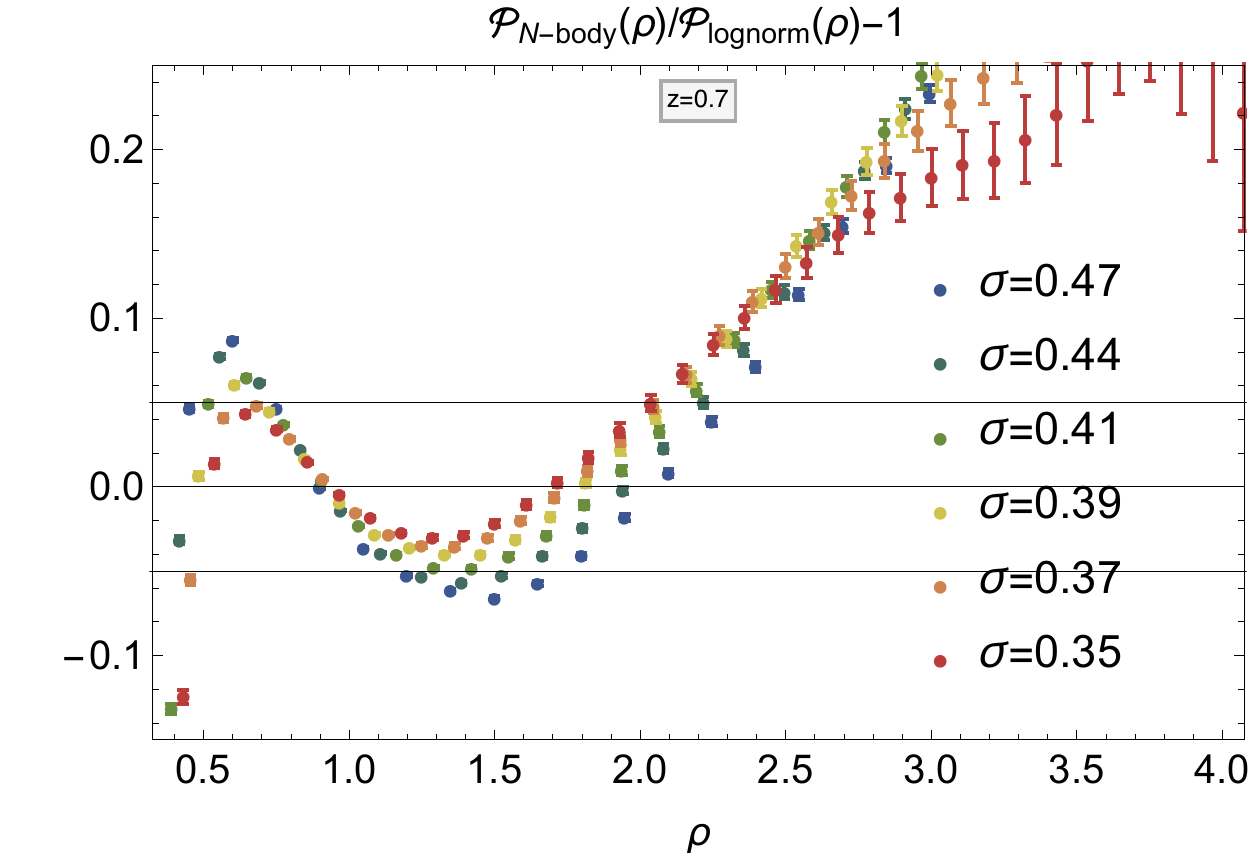}
   \caption{ 
   The residuals of the  best fit log normal PDF $\mP(\rho)$  for redshift $z=0.7$ and radii $R=10,11,...,15$Mpc/$h$ {\it (from blue to red)}  in comparison to the measurement from HR4 {\it (with error bars)}.
    This is to be contrasted to  the quasi perfect match of the saddle  PDF  presented in   Figure~\protect\ref{fig:compPDFsaddleHorizon}}
   \label{fig:lognormal-PDF-HR4-z0p7}
\end{figure}

\subsection{The density PDF in concentric spheres}
In Figure~\ref{fig:compPDFsaddleHorizon-z4}~and~\ref{fig:compPDFsaddleHorizon-z0} we show the one-cell PDFs for redshifts $z=0$ and $z=4$ comparing the saddle point approximation computed using {\tt LSSFast} with the measurements from the HR4 simulation. Furthermore we show the two-cell PDF for redshift $z=0$ in Figure~\ref{fig:compjointPDFsaddleHorizon-z0}. Note that we have chosen to use $\nu=1.59$ instead of $21/13\approx1.61$ at low redshift as the residuals were smaller in this case. This suggests that in order to get percent precision on the PDF at low redshift, one probably has to account for next-to-leading order correction to the skewness (that a slightly lower value of $\nu$ seems to reproduce). This is clearly seen at $z=0$ when the numerical integration of the inverse Laplace transform is carried out and shows residuals proportional to the typical third order Hermite polynomial, characteristic of the skewness. Adding higher order corrections to the skewness (by means of perturbation theory) is left for future work.

\begin{figure}
\includegraphics[width=0.99\columnwidth]{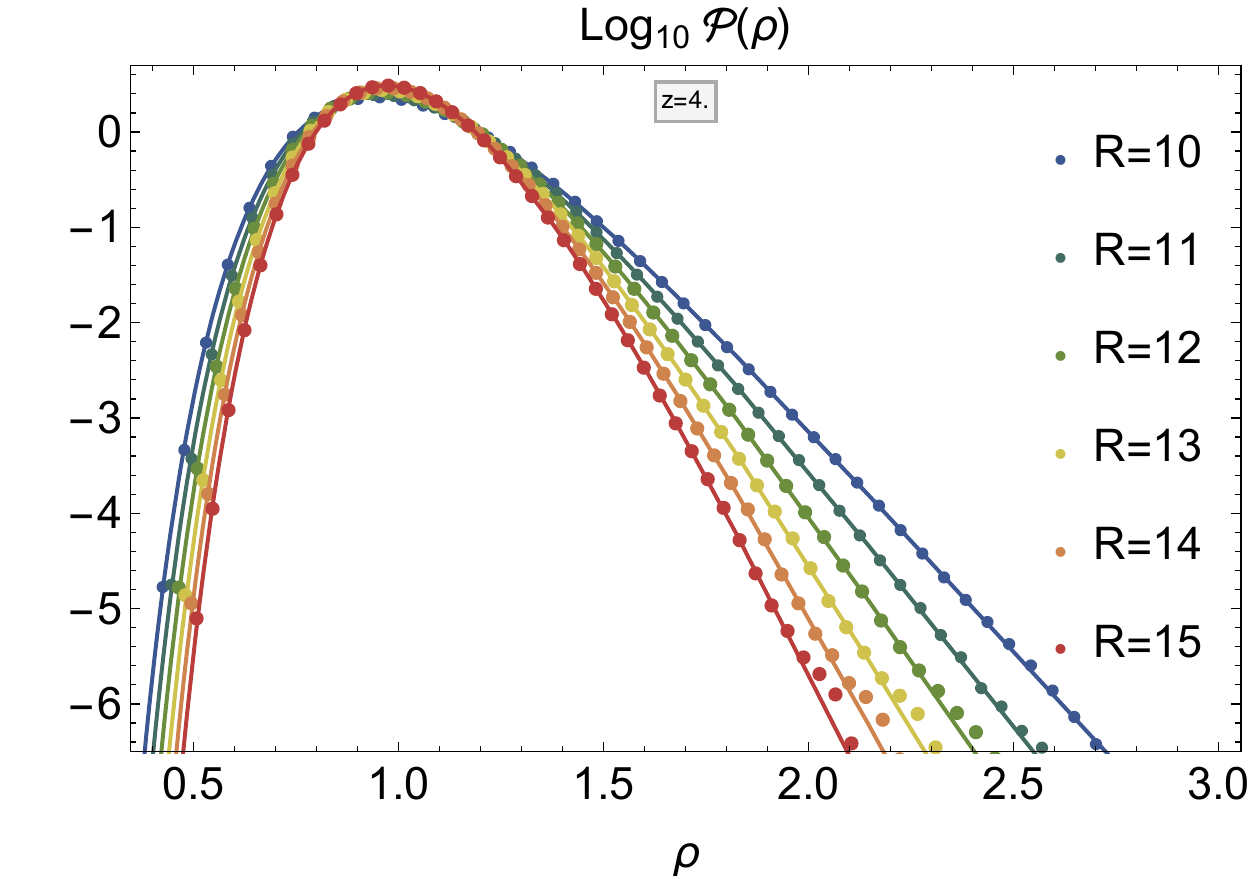}\vspace{0.2cm}\\
\includegraphics[width=0.99\columnwidth]{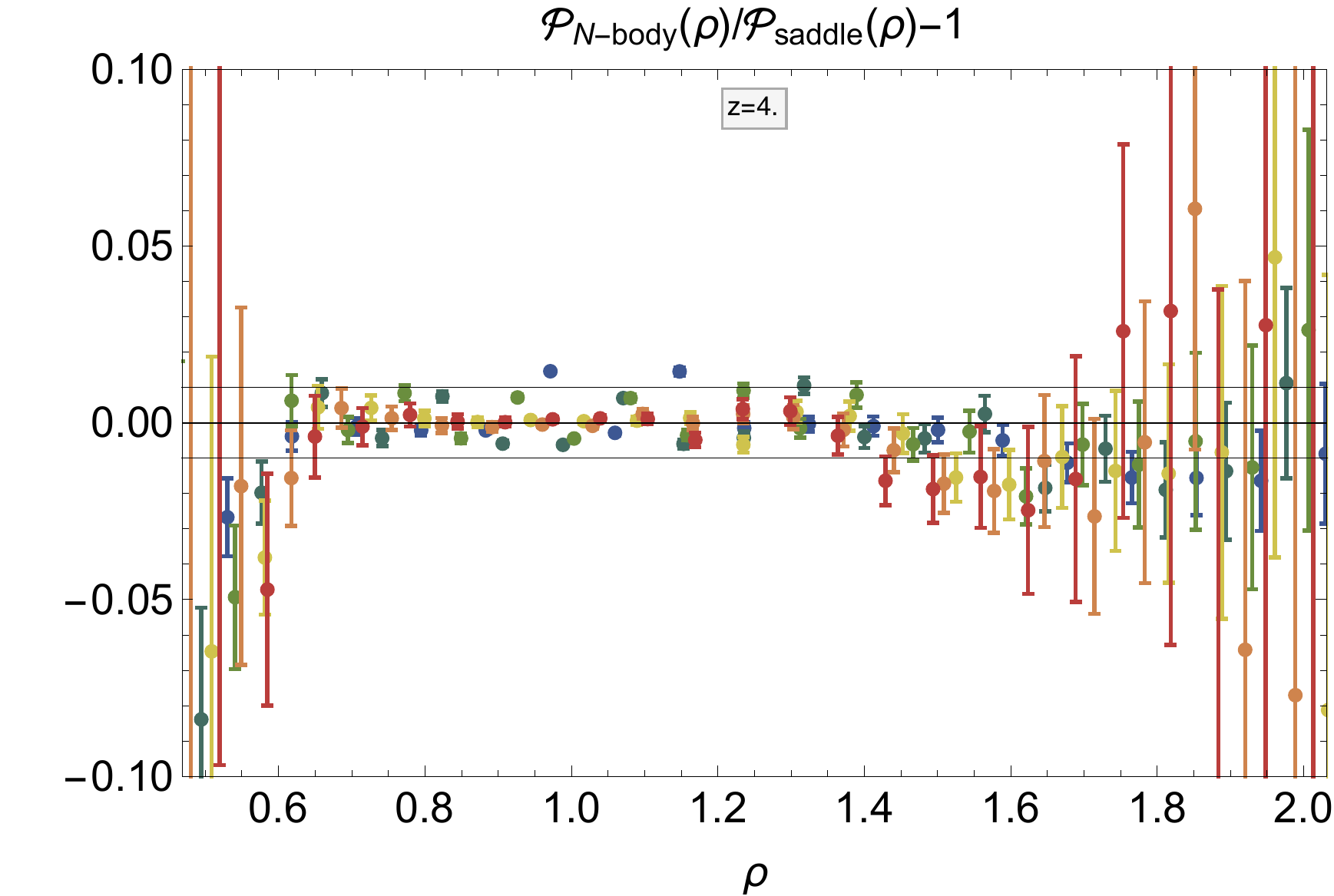}
   \caption{The density PDF $\mP(\rho)$ predicted from equation~\eqref{eq:saddlePDF1cell} using {\tt LSSFast} {\it (solid lines)} for radii $R=10,11,...,15$Mpc/$h$ {\it (from blue to red)} at redshift $z=4$ with $\nu=21/13$ in comparison to the HR4 measurement {\it (data points)} and the corresponding residuals.} 
   \label{fig:compPDFsaddleHorizon-z4}
\end{figure}

\begin{figure}
\includegraphics[width=0.99\columnwidth]{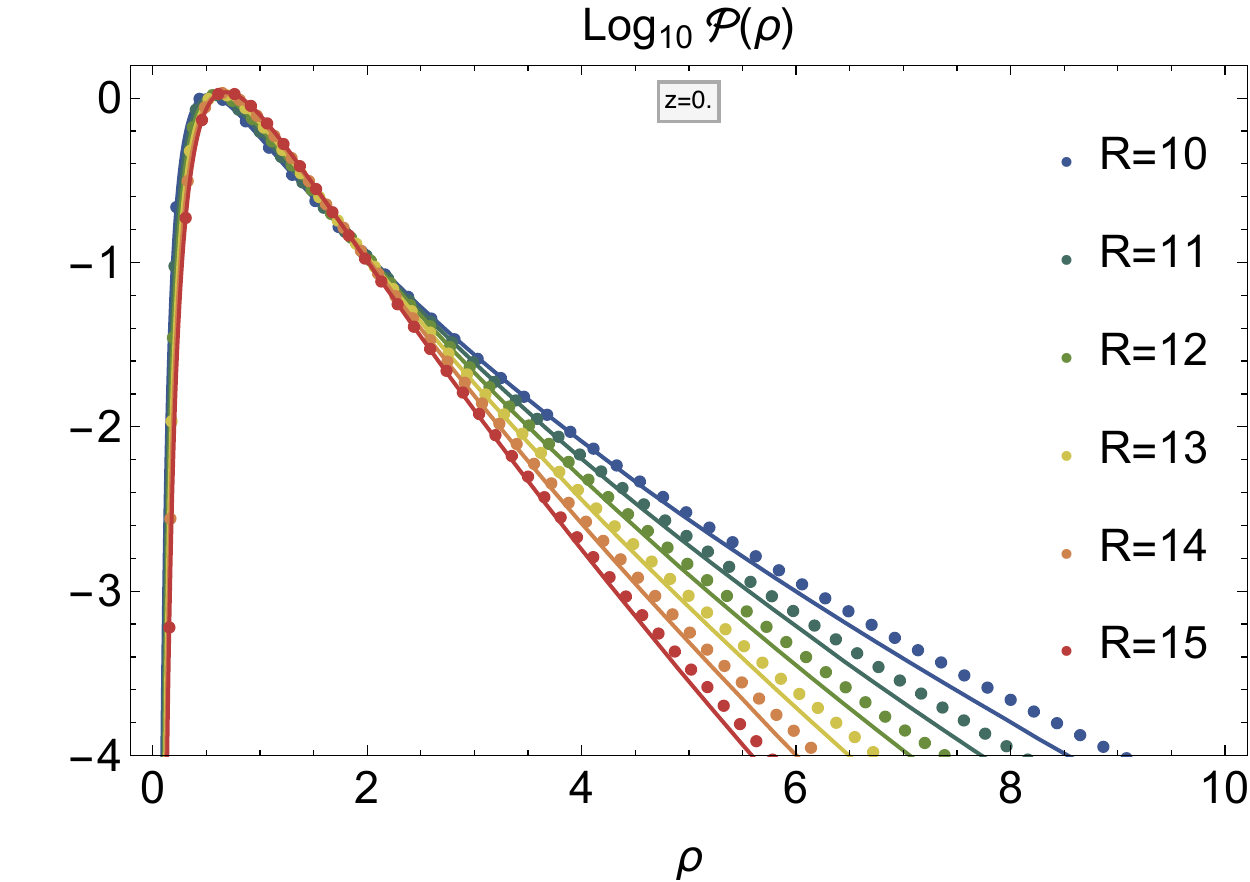}\vspace{0.2cm}\\
\includegraphics[width=0.99\columnwidth]{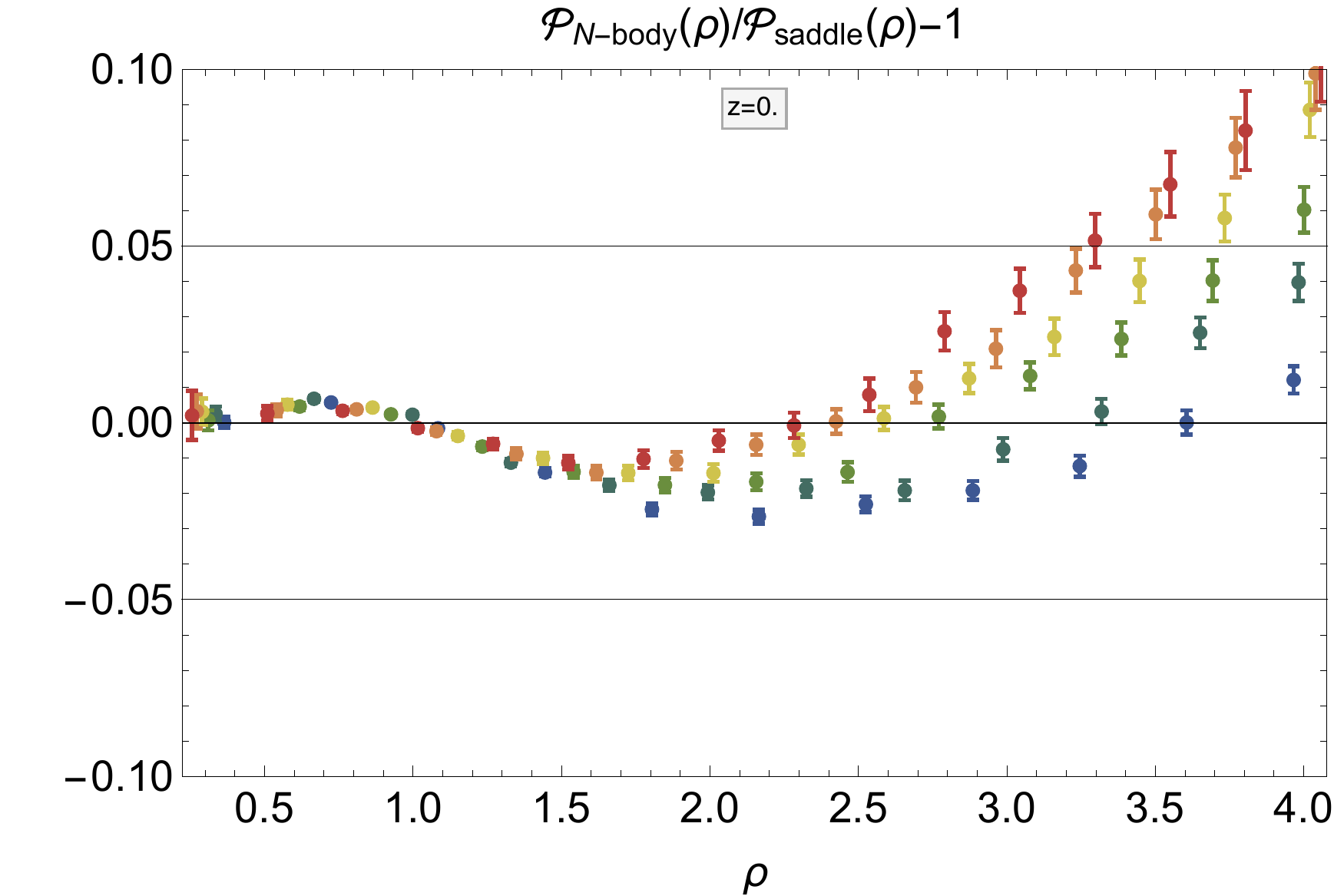}
   \caption{The density PDF $\mP(\rho)$ predicted from equation~\eqref{eq:saddlePDF1cell} using {\tt LSSFast} {\it (solid lines)} for radii $R=10,11,...,15$Mpc/$h$ {\it (from blue to red)} at redshift $z=0$ with a slightly adjusted $\nu=1.59$ in comparison to the HR4 measurement {\it (data points)} and the corresponding residuals.} 
   \label{fig:compPDFsaddleHorizon-z0}
\end{figure}

\begin{figure}
\centering
\includegraphics[width=0.95\columnwidth]{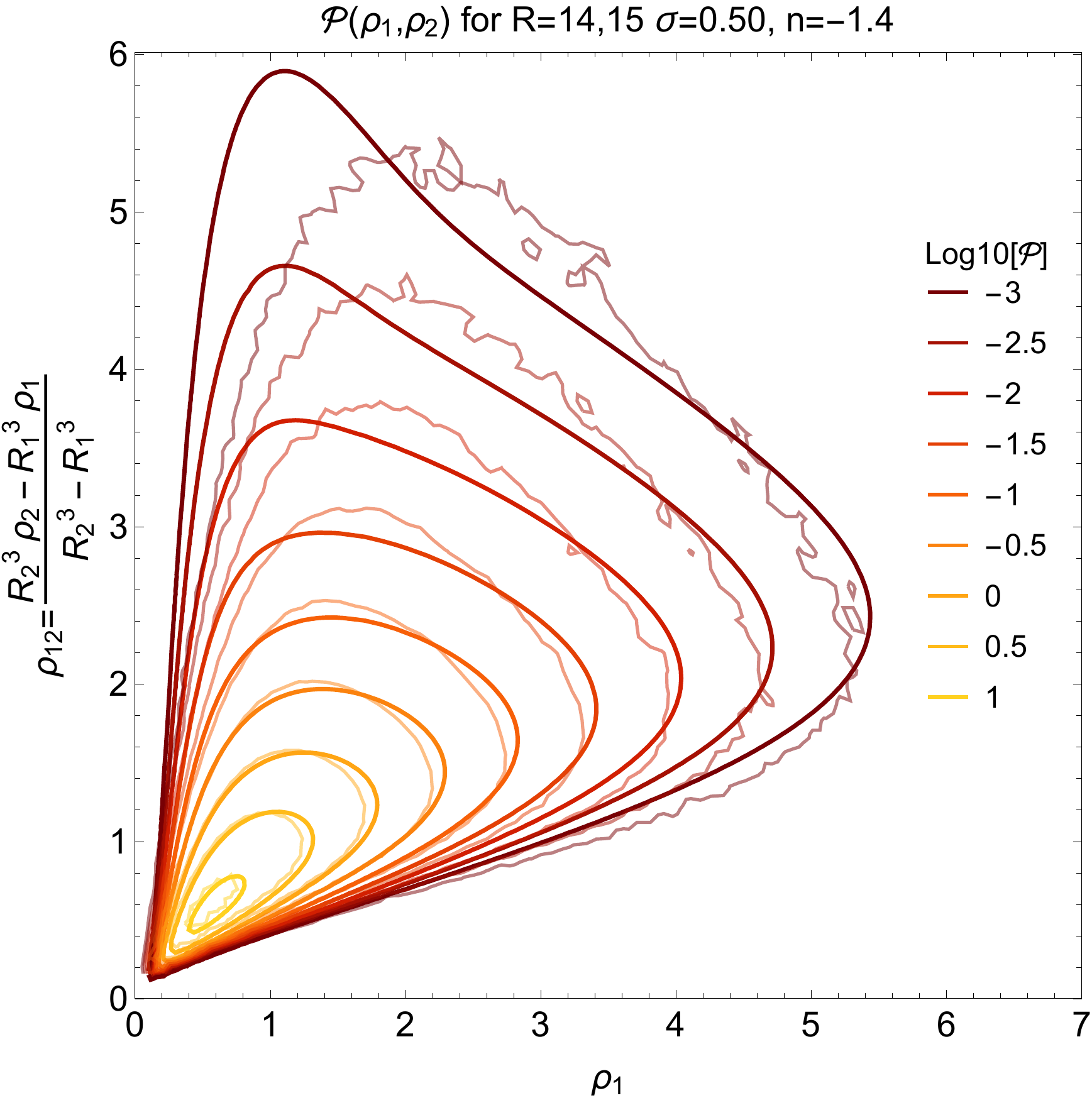}
   \caption{The joint density-slope PDF $\mP(\rho_1,\rho_2)$ predicted from equation~\eqref{eq:saddlePDF2cell} for radii $R_{1,2}=14, 15$ Mpc/$h$ at redshift $z=0$ with $\nu=21/13$ in comparison to the measurement from HR4 {\it (thin wiggly lines)}.} 
   \label{fig:compjointPDFsaddleHorizon-z0}
\end{figure}

\subsection{Bias functions}

\subsubsection{Density bias}
As a complement to Figure~\ref{fig:bias-rho-th} that shows the density bias for redshift $z=0.7$, we show in Figure~\ref{fig:bias-rho-th-z0} the corresponding result for redshift $z=0$ finding again excellent agreement with the simulation results.

\begin{figure}
\includegraphics[width=1.\columnwidth]{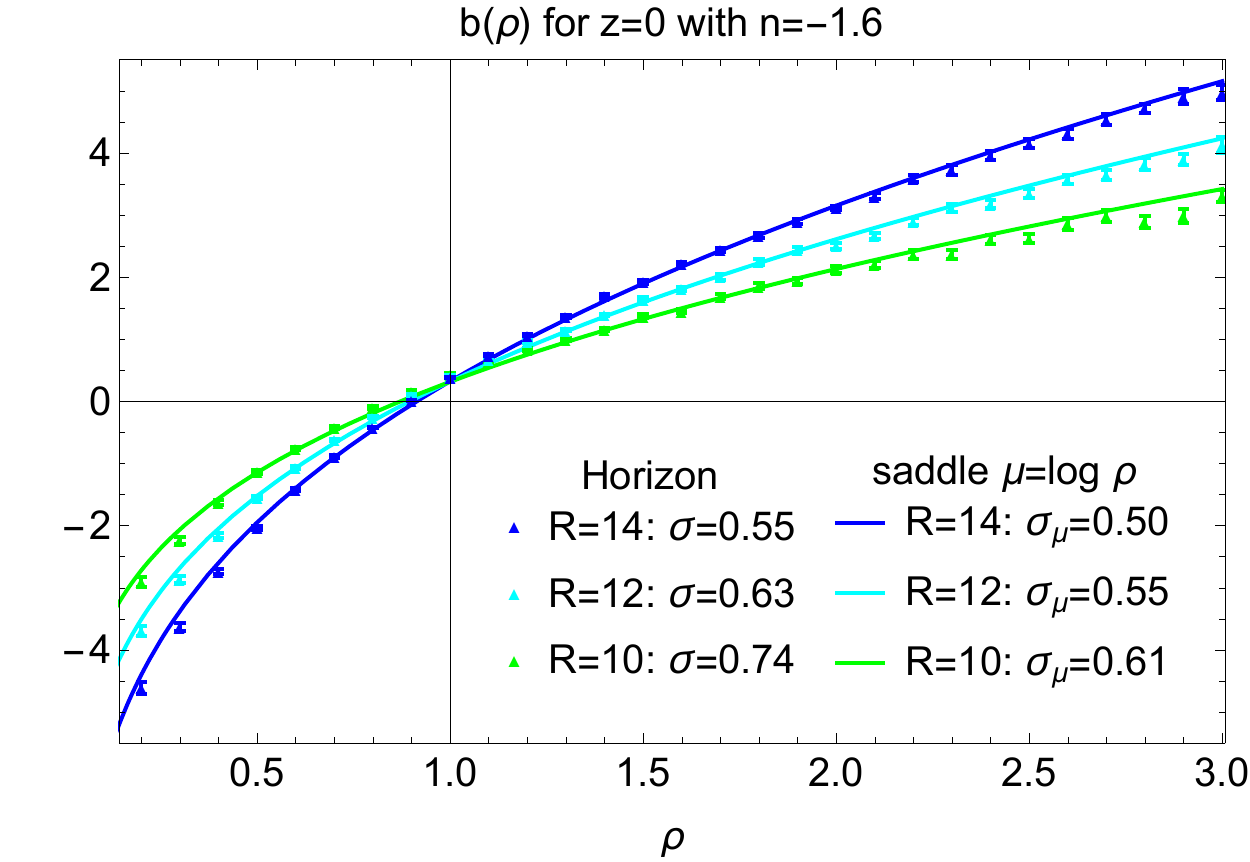}
   \caption{The density bias function $b(\rho)$ predicted from the saddle point approximation for the log-density $\mu=\log\rho$ at redshift $z=0$ for different radii and hence variances as indicated in the legend for $n_s=-1.6$.}
   \label{fig:bias-rho-th-z0}
\end{figure}

\subsubsection{Joint density slope bias}

As a complement to Figure~\ref{fig:bias-rho-slope-th} that shows the joint density bias for redshift $z=0.7$, we show in Figure~\ref{fig:bias-rho-slope-th-z0} the corresponding results for redshift $z=0$ finding again good agreement with the simulation.

\begin{figure}
\centering
\includegraphics[width=0.95\columnwidth]{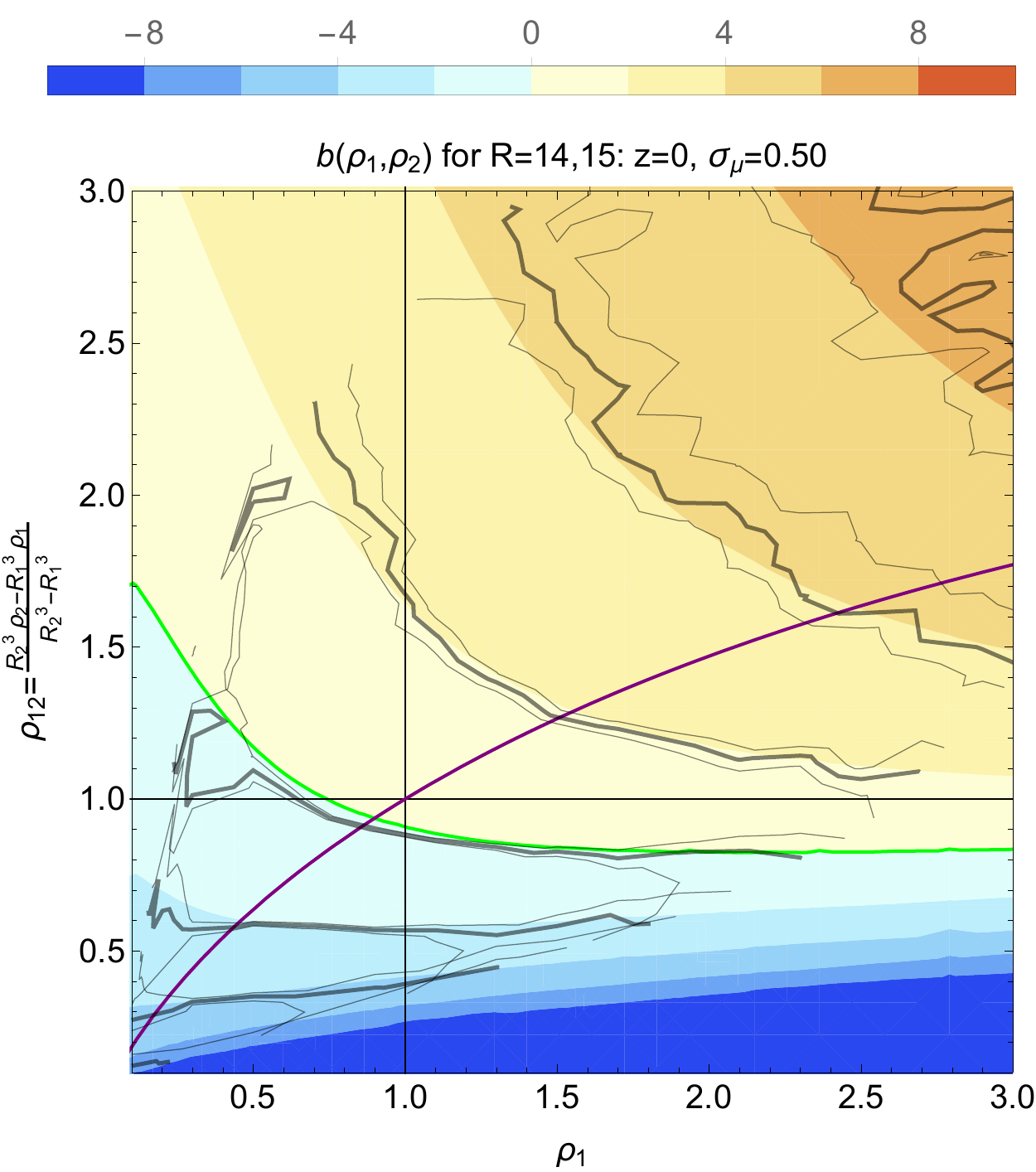}
   \caption{A contour plot of the joint density and slope bias function $b(\rho, s)$ predicted from the saddle point approximation equation~\eqref{eq:jointdensityslopebias} with normalization from equation~\eqref{eq:biasnorm} for $R_1=14$Mpc$/h$ and $R_2=15$Mpc$/h$ at redshift $z=0$ where $\sigma_\mu$=0.50, (corresponding to $\sigma_\rho=0.55$), in comparison to the measurements from HR4 (mean as thick black lines, and mean $\pm$ error on the mean as thin black lines).}
   \label{fig:bias-rho-slope-th-z0}
\end{figure}

\end{document}